\documentstyle[12pt,epsfig]{article}

\catcode`\@=11
\def\@citex[#1]#2{%
\if@filesw \immediate \write \@auxout {\string \citation {#2}}\fi
\@tempcntb\m@ne \let\@h@ld\relax \def\@citea{}%
\@cite{%
  \@for \@citeb:=#2\do {%
    \@ifundefined {b@\@citeb}%
      {\@h@ld\@citea\@tempcntb\m@ne{\bf ?}%
      \@warning {Citation `\@citeb ' on page \thepage \space 
undefined}}%
      {\@tempcnta\@tempcntb \advance\@tempcnta\@ne%
      \@tempcntb\number\csname b@\@citeb \endcsname \relax%
      \ifnum\@tempcnta=\@tempcntb 
	\ifx\@h@ld\relax%
	  \edef \@h@ld{\@citea\csname b@\@citeb\endcsname}%
	\else%
	  \edef\@h@ld{\ifmmode{-}\else--\fi\csname 
b@\@citeb\endcsname}%
	\fi%
      \else
	\@h@ld\@citea\csname b@\@citeb \endcsname%
	\let\@h@ld\relax%
      \fi}%
    \def\@citea{,\penalty\@highpenalty\,}%
  }\@h@ld
}{#1}}

\def\@citeb#1#2{{[#1]\if@tempswa , #2\fi}}
\def\@citeu#1#2{{$^{#1}$\if@tempswa , #2\fi }}
\def\@citep#1#2{{#1\if@tempswa , #2\fi}}

\def\bcites{         
	\catcode`\@=11
	\let\@cite=\@citeb
	\catcode`\@=12
}

\def\upcites{         
	\catcode`\@=11
	\let\@cite=\@citeu
	\catcode`\@=12
}

\def\plaincites{      
	\catcode`\@=11
	\let\@cite=\@citep
	\catcode`\@=12
}

\newcount\hour
\newcount\minute
\newtoks\amorpm
\hour=\time\divide\hour by 60
\minute=\time{\multiply\hour by 60 \global\advance\minute by-\hour}
\edef\standardtime{{\ifnum\hour<12 \global\amorpm={am}%
	\else\global\amorpm={pm}\advance\hour by-12 \fi
	\ifnum\hour=0 \hour=12 \fi
	\number\hour:\ifnum\minute<10 
0\fi\number\minute\the\amorpm}}
\edef\militarytime{\number\hour:\ifnum\minute<10 
0\fi\number\minute}

\def\draftlabel#1{{\@bsphack\if@filesw {\let\thepage\relax
   \xdef\@gtempa{\write\@auxout{\string
      \newlabel{#1}{{\@currentlabel}{\thepage}}}}}\@gtempa
   \if@nobreak \ifvmode\nobreak\fi\fi\fi\@esphack}
	\gdef\@eqnlabel{#1}}
\def\@eqnlabel{}
\def\@vacuum{}
\def\marginnote#1{}
\def\draftmarginnote#1{\marginpar{\raggedright\scriptsize\tt#1}}
\def\draft{
	\pagestyle{plain}
	\overfullrule=2pt
	\oddsidemargin -.5truein
	\def\@oddhead{\sl \phantom{\today\quad\militarytime} \hfil
	\smash{\Large\sl DRAFT} \hfil \today\quad\militarytime}
	\let\@evenhead\@oddhead
	\let\label=\draftlabel
	\let\marginnote=\draftmarginnote
	\def\ps@empty{\let\@mkboth\@gobbletwo
	\def\@oddfoot{\hfil \smash{\Large\sl DRAFT} \hfil}
	\let\@evenfoot\@oddhead}
	
\def\@eqnnum{(\theequation)\rlap{\kern\marginparsep\tt\@eqnlabel}%
	\global\let\@eqnlabel\@vacuum}  }

\def\blackfonts{
	\font\blackboard=msbm10 scaled\magstep1
	\font\blackboards=msbm8
	\font\blackboardss=msbm6
}

\def\nblack{            
	\def\ZZ{{Z \n{10} Z}}
	\def\NN{{N \n{14} N}}
	\def\CC{{C \n{11} C}}
	\def\RR{{R \n{11} R}}
	\def\QQ{{Q \n{12} Q}}
	\def\PP{{P \n{11} P}}
}

\def\prep{         
	\catcode`\@=11
	\input art10.sty
	\catcode`\@=12
	
	\let\small\null
	\def\blackfonts{
		\font\blackboard=msbm10
		\font\blackboards=msbm7
		\font\blackboardss=msbm5
	}
	\let\sl\it
	\twocolumn
	\sloppy
	\voffset=-2.54truecm
	\hoffset=-2.54truecm
	\flushbottom
	\parindent 1em
	\leftmargini 2em
	\leftmarginv .5em
	\leftmarginvi .5em
	\marginparwidth 48pt
	\marginparsep 10pt
	\setlength{\columnsep}{2truecm}
	\setlength{\textwidth}{25.4truecm}
	\setlength{\textheight}{17truecm}
	\baselineskip=16pt
	\oddsidemargin .18truein
	\evensidemargin .17truein
}

\def\eqalign#1{\null\,\vcenter{\openup\jot\m@th
  \ialign{\strut\hfil$\displaystyle{##}$&$\displaystyle{{}##}$\hfil
      \crcr#1\crcr}}\,}
\def\eqalignno#1{\displ@y \tabskip\centering
  \halign 
to\displaywidth{\hfil$\@lign\displaystyle{##}$\tabskip\z@skip
    &$\@lign\displaystyle{{}##}$\hfil\tabskip\centering
    &\llap{$\@lign##$}\tabskip\z@skip\crcr
    #1\crcr}}

\def\section{\@startsection {section}{1}{\z@}{3.ex plus 1ex minus
 .2ex}{2.ex plus .2ex}{\large\bf}}
\def\subsection{\@startsection{subsection}{2}{\z@}{2.75ex plus 1ex 
minus
 .2ex}{1.5ex plus .2ex}{\bf}}

\def\appendix{{\newpage\section*{Appendix}}\let\appendix\section%
	{\setcounter{section}{0}
	\gdef\thesection{\Alph{section}}}\section}

\def\abstract{\if@twocolumn
\section*{Abstract}
\else 
\begin{center}
{\bf Abstract\vspace{-.5em}\vspace{0pt}}
\end{center}
\quotation
\fi}

\catcode`\@=12

\newcommand{\beq}{\begin{equation}}
\newcommand{\eeq}{\end{equation}}
\newcommand{\beqa}{\begin{eqnarray}}
\newcommand{\eeqa}{\end{eqnarray}}

\newcommand{\Z}{{\bf Z}}

\newcommand{\C}{{\bf C}}
\newcommand{\e}{\,\,{\rm e}}

\def\noj#1,#2,{{\bf #1} (19#2)\ }
\def\jou#1,#2,#3,{{\sl #1\/ }{\bf #2} (19#3)\ }
\def\ann#1,#2,{{\sl Ann.\ Physics\/ }{\bf #1} (19#2)\ }
\def\cmp#1,#2,{{\sl Comm.\ Math.\ Phys.\/ }{\bf #1} (19#2)\ }
\def\ma#1,#2,{{\sl Math.\ Ann.\/ }{\bf #1} (19#2)\ }
\def\jd#1,#2,{{\sl J.\ Diff.\ Geom.\/ }{\bf #1} (19#2)\ }
\def\invm#1,#2,{{\sl Invent.\ Math.\/ }{\bf #1} (19#2)\ }
\def\cq#1,#2,{{\sl Class.\ Quantum Grav.\/ }{\bf #1} (19#2)\ }
\def\cqg#1,#2,{{\sl Class.\ Quantum Grav.\/ }{\bf #1} (19#2)\ }
\def\ijmp#1,#2,{{\sl Int.\ J.\ Mod.\ Phys.\/ }{\bf A#1} (19#2)\ }
\def\jmphy#1,#2,{{\sl J.\ Geom.\ Phys.\/ }{\bf #1} (19#2)\ }
\def\jams#1,#2,{{\sl J.\ Amer.\ Math.\ Soc.\/ }{\bf #1} (19#2)\ }
\def\grg#1,#2,{{\sl Gen.\ Rel.\ Grav.\/ }{\bf #1} (19#2)\ }
\def\mpl#1,#2,{{\sl Mod.\ Phys.\ Lett.\/ }{\bf A#1} (19#2)\ }
\def\nc#1,#2,{{\sl Nuovo Cim.\/ }{\bf #1} (19#2)\ }
\def\np#1,#2,{{\sl Nucl.\ Phys.\/ }{\bf B#1} (19#2)\ }
\def\pl#1,#2,{{\sl Phys.\ Lett.\/ }{\bf #1B} (19#2)\ }
\def\pla#1,#2,{{\sl Phys.\ Lett.\/ }{\bf #1A} (19#2)\ }
\def\pr#1,#2,{{\sl Phys.\ Rev.\/ }{\bf #1} (19#2)\ }
\def\prd#1,#2,{{\sl Phys.\ Rev.\/ }{\bf D#1} (19#2)\ }
\def\prl#1,#2,{{\sl Phys.\ Rev.\ Lett.\/ }{\bf #1} (19#2)\ }
\def\prp#1,#2,{{\sl Phys.\ Rept.\/ }{\bf #1C} (19#2)\ }
\def\ptp#1,#2,{{\sl Prog.\ Theor.\ Phys.\/ }{\bf #1} (19#2)\ }
\def\ptpsup#1,#2,{{\sl Prog.\ Theor.\ Phys.\/ Suppl.\/ }{\bf #1} 
(19#2)\ }
\def\rmp#1,#2,{{\sl Rev.\ Mod.\ Phys.\/ }{\bf #1} (19#2)\ }
\def\yadfiz#1,#2,#3[#4,#5]{{\sl Yad.\ Fiz.\/ }{\bf #1} (19#2) #3%
\ [{\sl Sov.\ J.\ Nucl.\ Phys.\/ }{\bf #4} (19#2) #5]}
\def\zh#1,#2,#3[#4,#5]{{\sl Zh.\ Exp.\ Theor.\ Fiz.\/ }{\bf #1} 
(19#2) #3%
\ [{\sl Sov.\ Phys.\ JETP\/ }{\bf #4} (19#2) #5]}

\def\eq#1{.~(\ref{#1})}

\def\beq{\begin{equation}}
\def\eeq{\end{equation}}
\def\beqar{\begin{eqnarray}}
\def\eeqar{\end{eqnarray}}

\newcommand{\tr}{{\rm Tr}}
\def\ssl#1{\rlap{\hbox{$\mskip 3 mu /$}}#1}     
\def\be{\begin{equation}}       \def\eq{\begin{equation}}
\def\ee{\end{equation}}         \def\eqe{\end{equation}}

\def\bea{\begin{eqnarray}}      \def\eqa{\begin{eqnarray}}
\def\ena{\end{eqnarray}}        \def\eea{\end{eqnarray}}
                                \def\eqae{\end{eqnarray}}
\def\nonu{\nonumber \\{}}

\def\Dbar{\bar{D}}
\def\al{\alpha}
\def\tr{{\rm Tr}}

\def\ppi{\Phi_i+\bar{\Phi}_i}

\def\nfrac#1#2{{\displaystyle{\vphantom1\smash{\lower.5ex\hbox{\sma
ll$#1$}}%
	\over\vphantom1\smash{\raise.25ex\hbox{\small$#2$}}}}}

\def\p#1{\mskip#1mu}
\def\n#1{\mskip-#1mu}
\def\stop{\p6.}
\def\comma{\p6,}

\def\to{\rightarrow}

\def\lae{\mathrel{\mathop{\smash{\lower .5 ex 
\hbox{$\stackrel<\sim$}}}}}
\def\lae{\mathrel{\mathop{\smash{\lower .5 ex 
\hbox{$\stackrel>\sim$}}}}}

\def\l:{\mathopen{:}\,}
\def\r:{\,\mathclose{:}}

\catcode`\@=11
\def\theequation{\arabic{equation}}
\catcode`\@=12

\nblack
\bcites

\nblack

\catcode`\@=11
\def\theequation{\thesection.\arabic{equation}}
\@addtoreset{equation}{section}
\@addtoreset{footnote}{section}
\@addtoreset{footnote}{subsection}
\catcode`\@=12

\newcommand{\var}{r}

\newcommand{\mr}{m_{r}}
\newcommand{\mc}{m_{c}}

\newcommand{\beqn}{\begin{equation}}
\newcommand{\eeqn}{\end{equation}}
\newcommand{\beqnarray}{\begin{eqnarray}}
\newcommand{\eeqnarray}{\end{eqnarray}}
\newcommand {\bear} [1] {\begin {array} {#1}}
\newcommand {\ear} {\end {array}}

\newcommand {\beqarn} {\begin{eqnarray*}}
\newcommand {\eeqarn} {\end{eqnarray*}}

\begin{document}
\begin{titlepage}

\begin{center}
\today
\hfill LBNL-40109, UCB-PTH-97/13\\
\hfill                  hep-th/9703100

\vskip 1.5 cm
{\large \bf Dynamics  of $N=2$ Supersymmetric
 Gauge Theories in
Three Dimensions}
\vskip 1 cm 
{Jan de Boer, Kentaro Hori and  Yaron Oz}\\
\vskip 0.5cm
{\sl Department of Physics,
University of California at Berkeley\\
366 Le\thinspace Conte Hall, Berkeley, CA 94720-7300, U.S.A.\\
and\\
Theoretical Physics Group, Mail Stop 50A--5101\\
Ernest Orlando Lawrence Berkeley National Laboratory, 
Berkeley, CA 94720, U.S.A.\\}

\end{center}

\vskip 0.5 cm
\begin{abstract}
We study the structure of the moduli spaces of vacua and
superpotentials of $N=2$ supersymmetric gauge theories in 
three dimensions. By analyzing the instanton corrections,
we compute the exact superpotentials and determine the
quantum Coulomb and Higgs branches of the theories in the
weak coupling regions. 
We find candidates for non-trivial
$N=2$ superconformal field theories at the singularities
of the moduli spaces. The analysis is carried out explicitly
for gauge groups $U(N_c)$ and $SU(N_c)$ with $N_f$ flavors.
We show that the field theory results are in complete
agreement with the intersecting branes picture. We also
compute the exact superpotentials for arbitrary gauge groups
and arbitrary matter content.

\end{abstract}

\end{titlepage}

\section{Introduction}

In the past, relatively little effort was directed towards the study of
$N=2$ supersymmetric gauge theories in three dimensions.
In \cite{AHW}, pure $N=2$ $SU(2)$ supersymmetric Yang-Mills theory
was studied and it was explained how instantons generate a superpotential
$e^{-\Phi}$ in the effective action. The non-perturbatively generated
superpotential for pure $SU(N_c)$ gauge theory can be found by
generalizing the results in \cite{AHW}. The explicit result was first
given in \cite{KV}, who used M-theory along the line suggested
in \cite{wi}.
It was rederived in \cite{dhoy} in the context of intersecting brane
configurations, along with results about mirror symmetry for $N=2$
theories.
The aim of this paper is to study the moduli spaces of vacua and 
the non-perturbatively
generated superpotentials of more general $N=2$ theories, 
in the weakly coupled regions.
The results can be extrapolated to the strong coupling regions of the moduli spaces
provided that the K\"ahler potential does not develop extra singularities. 

The paper is organized as follows:
In section 2 we introduce the basic properties of 
$N=2$ supersymmetric gauge theories in three dimensions.
These theories have both Coulomb and Higgs branches.
The former is parametrized by the vev's of the bosonic degrees of freedom
in the vector multiplet while the latter is parametrized by the vev's of the scalars 
in the
chiral multiplets.
Due to $N=2$ supersymmetry both the Higgs and the Coulomb branches are K\"ahler 
manifolds.
On the Coulomb branch the gauge group is generically broken to its maximal torus.
This is  in particular the case in the weakly coupled regions in which
our studies are concentrated.
In these regions the instanton calculus is reliable 
and we can study the non-perturbative
generation of superpotentials.

One of the important results
of section 2, which will be used in the later sections, is that 
the metric on the Coulomb branch
of $N=2$ abelian gauge theories is completely determined
at one loop. In order to show this
we use the superspace formalism. 
The detailed form of the metric and complex structure of the Coulomb 
branch are found by performing a duality transformation in superspace.
Another fundamental aspect of the structure that we find is that
the quantum Coulomb branch degenerates
at the points of massless electrons
and consists of several branches.
This can already be seen
for abelian $N=2$ gauge theories as discussed in that section.
However, the implications for the non-abelian cases are profound.
In that case, different superpotentials are generated on
different branches. 
In some branches, the superpotential may vanish.
Due to the singularities, none of this contradicts holomorphy. 

In section 3 we study the structure of the quantum Higgs and Coulomb branches and
the non-perturbative generation of a superpotential for the gauge group $SU(2)$
with $N_f$ quarks in the fundamental representation.
We show that the quantum Coulomb branch degenerates at the points of
massless quarks and consists of several branches.
We analyze in detail the non-perturbative generation
of a superpotential depending differently on different branches.
We find that a Higgs branch is not lifted by instanton corrections
and is emanating
from the singularity of the Coulomb branch, while
at the singularity itself the Lagrangian description
is not valid and the theory there is a candidate
for a non-trivial $N=2$ superconformal field theory.

In section 4 we study the structure of the quantum Higgs and Coulomb branches and
the non-perturbative generation of superpotentials for the gauge groups $U(N_c)$
and $SU(N_c)$
with $N_f$ quarks in the fundamental representation.
The structure that we find generalizes that of the previous section.
The quantum Coulomb branch develops singularitities separating
different regions. We analyze in detail the non-perturbative generation
of  superpotentials in the different
regions of the moduli space, as well as the structure of the Coulomb and Higgs
branches and their intersections.
We find many more candidates
for non-trivial $N=2$ superconformal field theories.
We review how the field theories can be obtained from an 
intersecting branes picture 
and show that the field theory results are in complete
agreement with string theory results for the intersecting branes.

In section 5 we generalize the results to arbitrary gauge group and arbitrary
matter content, and compute the exact superpotentials.

In the appendix we compute the number of fermionic zero modes of the $N=2$
theories in the background of monopoles which we use throughout the paper.

\section{$N=2$ gauge theories in three dimensions}

\subsection{$N=2$ supersymmetry}

 $N=2$ supersymmetric gauge theories in three 
dimensions can be obtained
by dimensionally reducing $N=1$ supersymmetric gauge theories in 
four 
dimensions. In this dimensional reduction the derivatives 
$\bar{D}_{\dot{\alpha}}$
and $D_{\alpha}$ become two complex conjugate derivatives 
$\bar{D}_{\alpha}$
and $D_{\alpha}$. To discuss the structure of the gauge multiplet 
it
is convenient to consider the algebra of covariant derivatives, and 
it is
here that there is a deviation between three and four dimensions. 
In three
dimensions the algebra can be chosen to be of the form\cite{hklr} 
\bea
\{\nabla_{\alpha},\nabla_{\beta}\}&  =  & 
\{\bar{\nabla}_{\alpha},\bar{\nabla}_{\beta} \} =0 
\nonu
\{\nabla_{\alpha} , \bar{\nabla}_{\beta} \} & = & i 
\nabla_{\alpha\beta} +
C_{\alpha\beta} G 
\eea
where $\nabla_{\alpha\beta}$ is the vector derivative and 
$C_{\alpha\beta}$
is the second Pauli matrix $\sigma_2$. The field-strength 
$G$ is a linear superfield
($\nabla^2 G = \bar{\nabla}^2 G = 0$), and its relation to the 
four-dimensional
field strength is roughly $W_{\alpha} = \bar{D}_{\alpha} G$. In the 
chiral
representation $\nabla_{\alpha}=e^{-V} D_{\alpha} 
e^V,\bar{\nabla}_{\alpha}=
\bar{D}_{\alpha}$, with $V$ an arbitrary Lie-algebra valued scalar 
superfield,
the field strength $G$ reads $G=\Dbar^{\al} (e^{-V} D_{\al} 
e^{V})$.
It transforms covariantly under the gauge transformations $e^V 
\rightarrow 
e^{i\bar{\Lambda}} e^V e^{-i \Lambda}$. 

The gauge kinetic term is now an integral over the full $N=2$ 
superspace,
\be
S_{kin} = \frac{1}{2e^2} \int d^3x d^2 \theta d^2 \bar{\theta} 
\tr(G^2).
\ee
Matter is included in the usual way, 
\be
S_{matter} = \int d^3 x d^2 \theta d^2 \bar{\theta} (Q^{\dagger} 
e^V Q + \tilde{Q}
e^{-V} \tilde{Q}^{\dagger}).
\ee
To avoid getting into issues of global ${\bf Z}_2$ anomalies and 
whether or not to
include Chern-Simons terms, we will always take pairs of chiral 
superfields 
$Q,\tilde{Q}$, and count such a pair as one quark multiplet. There 
are two different
kind of mass terms for quarks. There is a real mass term 
that can
only be written in three-dimensional $N=1$ superspace and reads
\be 
S_{m} = \int d^3 x d^2 \theta' (m_r Q^{\dagger} Q + \tilde{m}_r 
\tilde{Q}
 \tilde{Q}^{\dagger} ),
\ee
where $d^2 \theta'$ is the measure for $N=1$ superspace,
$d^2 \theta' = d^2(\theta+\bar{\theta})$. 
Again, to avoid anomalies, we will always take $m_r=-\tilde{m}_r$. 
Complex mass
terms can be obtained directly from four dimensions and read
\be
\int d^3 x d^2 \theta m_c \tilde{Q} Q + {\rm h.c.}
\ee
Finally, one may include Fayet-Iliopoulos terms 
$\zeta \int d^3 x d^2 \theta d^2 \bar{\theta} V$
and superpotential terms $\int d^3 x d^2 \theta P(Q,\tilde{Q})+ 
{\rm h.c.}$
in the action without destroying the $N=2$ supersymmetry.

\subsection{Low energy effective action}

The moduli space of vacua of $d=3,N=2$ gauge theories contains, in 
general,
a Coulomb and an Higgs branch. By $N=2$ supersymmetry, both these 
moduli
spaces are K\"ahler manifolds, but neither one of them is protected 
by
some non-renormalization theorem against loop or non-perturbative
corrections. The low-energy effective action is an $N=2$ 
supersymmetric
sigma model that can be written as 
\be \label{jjj1}
S=\int d^3 x d^2 \theta d^2 \bar{\theta} K(\Phi_i,\bar{\Phi}_i) +
 \int d^3 x d^2 \theta W(\Phi_i) + {\rm h.c.},
\ee
where $K$ is the K\"ahler potential on the moduli space and 
$W(\Phi_i)$
some superpotential. On the Higgs branch, the $\Phi_i$ are suitable
gauge invariant combinations of the matter fields $Q,\tilde{Q}$. 
On the Coulomb 
branch, at a generic point, the gauge group is broken to a product
of $U(1)$'s. If we denote by $\varphi$ the scalar and by $A_{\mu}$ the
vector in the $N=2$ vector multiplet, then the action contains
terms $\tr[A_{\mu},A_{\nu}]^2$ and $\tr[A_{\mu},\varphi]^2$. These 
should
be zero for a supersymmetric vacuum, which shows that we can choose
both $\varphi$ and $A_{\mu}$ to lie in a Cartan subalgebra of the 
gauge
group. 
Part of the coordinates that parametrize the Coulomb branch
are the vacuum expectation values $r_i=<\varphi_i>=<G_i>$.  
The other half of the coordinates
are provided by the vacuum expectation values of scalar fields that
one obtains by dualizing the gauge field in three dimensions. 
We will denote the scalar fields dual to the gauge fields by
$\sigma_i$, and use the same symbol for its expectation value. The
two real scalars $\varphi$ and $\sigma_i$ are related to the bosonic 
components
of the chiral superfield $\Phi_i$ in (\ref{jjj1}). In the next 
section
we will show explicitly how one performs the duality transformation
to arrive at an action of the type (\ref{jjj1}), starting from the 
low-energy
effective action in terms of the massless vector superfields $V_i$. 
 
The Wilsonian effective action for the massless degrees of freedom 
on
the Coulomb branch is obtained by integrating out all massive 
degrees
of freedom, i.e. all charged matter multiplets and the off-diagonal
components of the vector multiplets. The resulting effective action 
will
be a functional of the remaining massless vector multiplets $V_i$, 
and in view
of the gauge invariance of
the effective action it will be a function of the field
strengths $G_i$ of $V_i$ only. Since we are interested in the 
low-energy effective
action, we will discard all terms in the action that contain 
derivatives of
$G_i$ and keep only the part that is purely algebraic in $G_i$,
\be
S_{{\rm low~ energy}} = \int d^3 x d^2 \theta d^2 \bar{\theta} 
f(G_i).
\ee
In general we expect arbitrary higher loop corrections in the 
function
$f$. There are two important exceptions. First, if we are 
considering a theory
with $N=4$ supersymmetry, we expect only a one-loop correction to 
$f$, but
no higher loop corrections, although it may receive 
non-perturbative 
corrections. The second case is when the original gauge theory is 
abelian.
In that case the action is bilinear in the massive fields, and the 
effective
action has only a one-loop contribution. Furthermore, as there are 
no
monopoles in an abelian gauge theory, nonperturbative corrections 
will
be absent and the one-loop result is exact. 

A similar analysis applies when there are neutral chiral multiplets $M$ 
in the theory.
In that case, the low energy effective action will also include 
some of
their degrees of freedom (which ones depends on the precise form of 
the
superpotential), and more generally we have
\be \label{j1}
S_{{\rm low~ energy}} = \int d^3 x d^2 \theta d^2 \bar{\theta} 
f(G_i,M,\bar{M}).
\ee
Rather than computing the one-loop corrections to $f$, we will find 
it more
convenient to first perform a duality transformation. The one-loop
corrections in the dual variables are already known 
\cite{intsei,dhoo}, and
by performing an inverse duality transformation one may easily 
obtain the
one-loop corrections to $f$. 

\subsection{Duality transformation}

To perform the duality transformation, consider \cite{hklr}
\be \label{j3}
S = \int d^3 x d^2 \theta d^2 \bar{\theta} 
(f(\tilde{G}_i,M,\bar{M})-
 \tilde{G}_i(\ppi)),
\ee
where the $\tilde{G}_i$ are unconstrained real superfields,
and the $\Phi_i$ are chiral superfields. By varying $\Phi_i$ we
find that $\tilde{G}_i$ is a linear superfield
and substituting this in the action we
reobtain (\ref{j1}). Another possibility is to vary the action with 
respect
to $\tilde{G}_i$. This leads to the equation
\be \label{j2}
\ppi =  \frac{\partial f(\tilde{G}_i,M,\bar{M}) }{\partial 
\tilde{G}_i}.
\ee
By solving this for $\tilde{G}_i$ in terms of $\ppi$, and 
substituting this back into
the action (\ref{j3}), we obtain the dual description
\be \label{j4}
S =  \int d^3 x d^2 \theta d^2 \bar{\theta} K(\ppi,M,\bar{M}),
\ee
where $K$ is the Legendre transform of $f$. The action (\ref{j4}) 
is 
the standard action for a set of chiral superfields, and the moduli
space is the complex manifold parametrized by the vevs of $\Phi_i$
with K\"ahler potential $K$. Even if $f$ has only one-loop 
corrections,
$K$ will have corrections to all orders in the coupling constant,
due to the fact that one has to invert the relation (\ref{j2}).

The duality transformation, when worked out in components, contains
the usual duality transformation to go from a gauge field to a 
scalar $\sigma$. This scalar is the imaginary scalar component of 
the superfield $\Phi$. As one can see by considering Wilson lines
for the gauge field, the scalar $\sigma$ should be compact. With 
the
normalizations above, this means that we have to identify
\be
\Phi \sim \Phi + 2\pi i \lambda_k
\ee
for each fundamental weight $\lambda_k$ of the Lie algebra of the 
gauge 
group. In other words,
$\Phi$ is a multivalued coordinate on the moduli space, and the
good single valued superfields are $e^{\alpha_i \cdot \Phi}$, where
$\alpha_i$ is a simple root. These superfields
also provide the natural complex structure on the moduli space.
This already severely constrains the form of a superpotential:
it must be a holomorphic function of $e^{\alpha_i \cdot \Phi}$.

The action (\ref{j4}) is obviously invariant under constant shifts
of $\sigma$. In view of the compactness of $\sigma$, these constant
shifts form ${\rm rank}(G)$ commuting holomorphic $U(1)$ isometries
of the metric on the moduli space. These symmetries cannot be 
broken
in perturbation theory, but can be broken by a non-perturbatively
generated superpotential.

Finally it is important to notice that due to the identity 
(\ref{j2})
the relation between $<\Phi_i>$ and $r_i \equiv <G_i>$ can be
quite complicated. This will play a crucial role later on in
order that the non-perturbative superpotential that we find 
is not in conflict with results from index theory, but does
in fact confirm these results. It will turn out that the chiral
superfields $\Phi_i$ are not always good coordinates on the
Coulomb branch. Their real part may become $-\infty$, and if
this happens we have to consider different coordinate patches
to discuss the Coulomb branch. This allows for the possibility
to have different superpotentials on different pieces of the 
Coulomb branch, without violating holomorphy. In the sequel,
we will see how this will enable us to arrive at a consistent
picture for the dynamics on the Coulomb branch of $N=2$
theories with arbitrary matter.

\subsection{Example: $U(1)$ with one flavor}

The one-loop metrics described in \cite{intsei,dhoo} were expressed 
in
the `mixed' variables $r_i=<G_i>$ and $\sigma_i={\rm Im} 
(<\Phi_i>)$.
To illustrate how one derives such a metric in the above picture we
will discuss as an example the case of a $U(1)$ gauge theory with
one electron. The one-loop metric in this case reads
\be \label{j9}
ds^2 = \frac{1}{4} \left( \frac{1}{e^2} + \frac{1}{|r|} \right) 
dr^2 + 
\left( \frac{1}{e^2} + \frac{1}{|r|} \right)^{-1} d\sigma^2
\ee
Classically, the moduli space is a cylinder, but quantum 
mechanically
it is broken up in two pieces ($r>0$ and $r<0$), that both look
like a rounded-off half-cylinder. Both half-cylinders touch at a
point. At this point we can neglect $1/e^2$ and the metric looks
like that of two complex planes touching at the origin, $ds^2
\sim dz d\bar{z}$ with $z=\sqrt{|r|}e^{i\sigma}$. The difference
between the tree-level and one-loop metric is illustrated in
figure 1.  

\begin{figure}
\begin{center}
$\mbox{\epsfig{figure=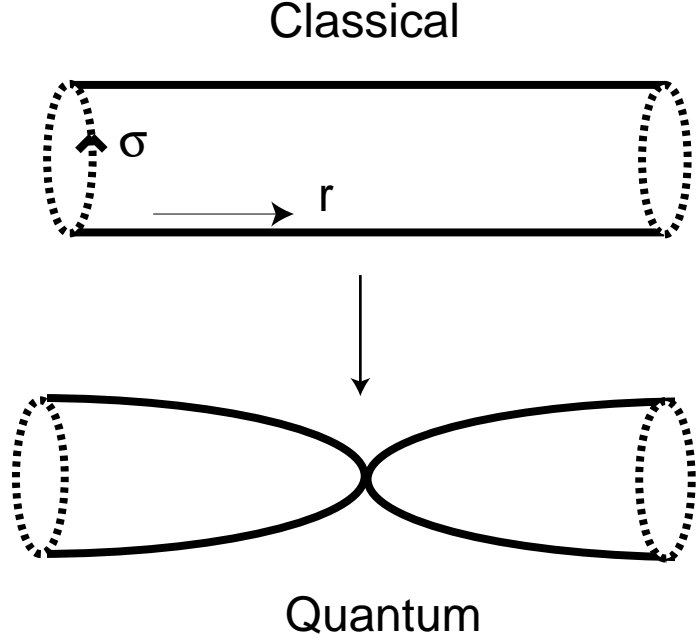}}$
\end{center}
\caption{The classical and quantum moduli spaces of vacua of $U(1)$ 
theory.}
\end{figure}

The one-loop low-energy effective action for this theory reads
\be \label{j5}
f(G)=\frac{1}{2e^2} G^2 + G \log (G/e^2)
\ee
and from (\ref{j2}) we obtain the relation between $\Phi$ and $G$,
\be \label{j6}
\Phi=\frac{1}{2e^2} G + \frac{1}{2}\log (G/e^2) +\frac{1}{2} + 
i\sigma.
\ee
The sigma-model metric of the dualized theory reads
\be \label{j7}
ds^2 = K''(\Phi+\bar{\Phi}) d\Phi d\bar{\Phi}
\ee
where we denote by $\Phi$ both the superfield and its vev.
To express $ds^2$ in terms of $r=<G>$ and $\sigma$ we use the
fact that the inverse of a Legendre transform is again a Legendre
transform, and therefore
\be K'(\Phi+\bar{\Phi})=r. \ee
Differentiating once more, we find 
\be \label{j8}
 K''(\Phi+\bar{\Phi}) = \left( \frac{\partial 
(\Phi+\bar{\Phi})}{\partial r} 
\right)^{-1}=\left( \frac{1}{e^2} + \frac{1}{r} \right)^{-1}.
\ee
Thus, for the metric we finally get
\be
ds^2 =\left( \frac{1}{e^2} + \frac{1}{r} \right)^{-1} \left(
\frac{1}{2} \left( \frac{1}{e^2} + \frac{1}{r} \right)dr + id\sigma 
\right) \left(
\frac{1}{2} \left( \frac{1}{e^2} + \frac{1}{r} \right)dr - id\sigma 
\right),
\label{abel}
\ee
which indeed agrees with (\ref{j9}). This is the exact result,
and by expanding it in powers of $e^2$ one can see that it
does contain terms with arbitrary high powers of $e^2$. However,
beyond one-loop they are all a result of the duality 
transformation.

In this simple case we already see the phenomenon that the chiral
superfield $\Phi$ is not a good coordinate everywhere on the moduli
space. Its real part becomes $-\infty$ as we approach the point 
$r=0$
where the electron becomes massless, and the moduli space
decomposes in two different parts, only half of which is described 
by
$\Phi$. A priori, there can be different non-perturbative dynamics
on each part of the moduli space, without violating holomorphy, and 
we
will see in examples of non-abelian gauge theories that this is 
indeed
what happens.

\subsection{General abelian gauge theories}

We will now discuss the case of a general abelian gauge theory, 
with
gauge group $U(1)^{N_c}$, and $N_f$ electrons $Q_{a},
\widetilde{Q}_{a}$, with charges $q_{ia}$ under the $i$-th
$U(1)$ gauge group. In addition, we will assume that the 
electrons
have a real mass $m_{r}^{a}$ and that there are superpotential
terms $\int d^3 x d^2 \theta m_{c}^{a} \widetilde{Q}_{a} 
Q_{a}$.
The metric in terms of $r_i$ and
$\sigma_i$, $i=1\ldots N_c$ can be extracted from the
results in \cite{dhooy}.
We will be mainly interested in the relation between $\Phi_i$ and 
$r_i$,
and it is this relation that we will compute here. According to the 
discussion
above, the second derivative of $K$ yields directly the coefficient 
in
front of $d\sigma_i d\sigma_j$ in the metric on the moduli space. 
On the other hand, the second derivative of $K$ is proportional
to $\partial r_j /\partial (\ppi)$. Combining these observations 
with
the results of \cite{dhooy} we obtain
\be \label{j10}
 \frac{\partial(\ppi)}{\partial r_j} = \frac{\delta_{ij}}{e^2} + 
\sum_{a} \frac{q_{ia}
q_{ja}}{\sqrt{(\sum_k q_{ka} r_k - m_r^{a})^2 + 
|m_c^{a}|^2}}.
\ee
This can be integrated to yield
\be \label{j11}
\Phi_i = \frac{r_i}{2e^2} + \frac{1}{2} \sum_{a} q_{ia} 
\log \left( \sum_k q_{ka} r_k - m_r^{a} + 
\sqrt{(\sum_k q_{ka} r_k - m_r^{a})^2 + |m_c^{a}|^2} 
\right) + i\sigma_i.
\label{one}
\ee
The one-loop result for a general non-abelian gauge theory has
a similar form, with an extra term 
\be 
-\frac{1}{4} \sum_{b} q_{ib} \log \left(\sum_k q_{kb} r_k
\right)
\ee
in the right hand side of (\ref{j11}), where $q_{ib}$ are the 
$U(1)$
charges of the various massive vector bosons labeled by $b$.

\subsection{Preliminary remarks on non-abelian gauge theories}

In the following sections, we analyze non-abelian gauge theories.
The essential difference between abelian and non-abelian gauge theories
is that non-perturbative effects are present
in non-abelian gauge theories because of instanton
configurations, which are BPS monopoles from the four-dimensional
point of view. In particular, a superpotential can be generated 
only by such non-perturbative effects. To determine whether or not
a particular monopole configuration can contribute to the superpotential
we use zero mode counting and 
the $U(1)_R$ symmetry. The number of fermionic zero modes in a
monopole background is determined by the index of the corresponding
Dirac operator. This index is computed in the Appendix. The results for
 $SU(N_c)$ gauge theory with $N_f$ quark
multiplets, i.e., $N_f$ pairs of chiral multiplets in the fundamental
and the dual representation, are as follows.
Let us consider a magnetic monopole
in the unbroken gauge group $U(1)^{N_c-1}$
of magnetic
charge $(n_1,n_2,\ldots,n_{N_c-1})$. In this configuration,
there are $2\sum n_i$ zero modes from gluinos.
The number of zero modes from quark multiplets depends on the vev of
the scalar field $\varphi={\rm diag}(\var_1,\ldots ,\var_{N_c})$.
By using the Weyl group action, we may assume that
$\var_{i}>\var_{i+1}$.
Let us focus on the $a$-th quark. If its real bare mass
is in the interval $\var_{i}>\mr^a >\var_{i+1}$ for some $i$,
it has $2n_i$ zero modes. If its real bare mass is larger than $\var_1$
or smaller than $\var_{N_c}$, there are no zero modes. We will not 
discuss the other
cases in which some $\var_i$ coincide with $\mr^a$.
Since gluinos have $U(1)_R$ charge $+1$ and fermions in quark multiplets
have charge $-1$, a superpotential can be generated by instantons in this
sector only when
\beq
2\sum_{i=1}^{N_c-1} n_i-2\sum_{a=1}^{N_f}\ell_a=2
\eeq
where $\ell_a:=n_i$ if $\var_{i}>\mr^a>\var_{i+1}$ and $\ell_a=0$ if
$\mr^a>\var_1$ or $\mr^a<\var_{N_c}$. If there is no complex mass and
the quark vevs all vanish there are additional global $U(1)$ symmetries.
In particular, a combination of $U(1)_R$ and the axial $U(1)$ flavor
symmetry leads to the additional selection rule 
\beq \label{selec2}
2\sum_{i=1}^{N_c-1} n_i+2\sum_{a=1}^{N_f}\ell_a=2.
\eeq

An interesting consequence of the above
formula is that instanton sectors that may
possibly contribute to the superpotential can
change as $\varphi$ is varied, e.g., as we move
from $\var_i>\mr^a$ to $\var_i<\mr^a$.
This suggests that the superpotential can jump along real codimension one
hypersurfaces $\var_i=\mr^a$ of the Coulomb branch,
which seems peculiar in view of the fact that the
superpotential should be a holomorphic function of
the chiral multiplets. The resolution of this as  we will observe
in the following sections
is that the Coulomb branch degenerates
along the hypersurfaces which actually have complex codimension one
and consists of several branches, on each of which the
superpotential is holomorphic. This degeneration is
caused by loop corrections to the metric. 
In fact, near the ``phase boundary''
$\var_i=\mr^a$, the theory can be well approximated by
the $U(1)$ gauge theory with a single electron (tensored with
the remaining free $U(1)^{N_c-2}$ maxwell theory),
and we know the behavior
of the Coulomb branch of such a theory near the point of a 
massless electron.
As discussed in section 2.4,
it is exactly determined at the one-loop level, and the structure
of degeneration is depicted in Figure 1. 

In the following sections we will always consider the weakly coupled region
of the moduli space, i.e. the region where $|r_i-r_j|\gg e^2$. In
this region, the off-diagonal components of the vector multiplet
are very heavy and the theory can be approximated by an abelian
gauge theory with $N_cN_f$ electrons. The abelian theory
has only one-loop corrections, and this is the reason that the 
one-loop results are an accurate description of the moduli
space at weak coupling. 
The results can be extrapolated to the strong coupling regions
provided we assume that 
the K\"ahler potential does not develop extra singularities there.

\section{Dynamics of $SU(2)$ Gauge Theories}

In this section, we determine the structure of vacua in the weak
coupling region of
$SU(2)$ gauge theories with matter chiral multiplets
in the fundamental representation. In particular,
we will see how
the classical Coulomb branch degenerates
to have several branches due to one loop effects and
how one of them is lifted due to
the non-perturbative generation of a superpotential.

The classical Coulomb branch is parametrized by the vevs of
the scalar field $\var$ of the vector multiplet
in the unbroken $U(1)$ subgroup
and the scalar $\sigma$ dual to the $U(1)$ vector field.
The classical metric looks like
\beq
ds^2={1\over 4e^2}d\var^2+e^2d\sigma^2.
\label{classmet}
\eeq
It is the cylinder $\{(\var,\sigma)\}$;
$\sigma\equiv \sigma+2\pi$ divided by the action of the
Weyl goup $(\var,\sigma)\mapsto (-\var,-\sigma)$.
As we increase the vev $|\var|$,
loop and instanton corrections to (\ref{classmet}) and
other physical quantities decrease
as inverse powers of $\var/e^2$ and
exponentially $\e^{-|\var|/e^2}$ respectively. 
Conversely, if we decrease $|\var|$ below $e^2$
we expect a strong dynamics involving gluons and gluinos.
For instance, the one-loop corrected metric
of the pure Yang-Mills theory is
$ds^2=gd\var^2/4+g^{-1}d\sigma^2$ where
\beq
g={1\over e^2}-{2\over |\var|}\,.
\eeq
We observe that below $|\var|=2e^2$
the one-loop metric becomes negative
definite and higher loop and instanton
corrections are needed to render the metric
positive definite.
This is the analog of the QCD scale
$\Lambda$ in four-dimensional asymptotically free gauge
theories where the one-loop gauge coupling constant
diverges.
In what follows, we focus on the weak coupling region
$|\var|\gg e^2$
in which the semi-classical approximation is valid.
By using the Weyl group action,
we can restrict attention to the region $\var\gg e^2$.
We also assume that all the real bare masses of the quarks
are much greater than $e^2$ so that there are no massless
quarks in the strong coupling region.

A superpotential can be generated only by non-perturbative
effects. 
Since we are considering the region $\var \gg e^2$ where
we do not expect strong dynamical effect,
a superpotential can be generated only by instantons,
which are the magnetic monopoles in the unbroken $U(1)$
subgroup of $SU(2)$.
A superpotential is generated only when a selection rule
coming from a conserved fermion number symmetry $U(1)_R$
is satisfied.
Under this $U(1)_R$ global symmetry,
the ``gluinos'' (fermions in the $SU(2)$ vector multiplet)
have charge $+1$ and the fermions in the matter chiral
multiplet have charge $-1$, and a possible F-term carries
charge $2$.

To see whether the superpotential is generated, we need
to count the number of fermion zero modes of Dirac operator
in the instanton configuration. The results are summarized in
section~2.6.
In the configuration of magnetic charge $n$, there are
$2n$ zero modes from gluinos. The number of fermion
zero modes in the matter chiral multiplet depends
on the vev $\var$. From each quark multiplet,
there are $2n$ zero modes if $\var$ is larger than
the real bare mass $\mr$, while there are no zero modes
if $\var < \mr$.
Thus, the selection rule is now expressed as
\beq
\label{au1}
2n-2nk=2\,,
\eeq
where $k$ is the number of quarks whose real bare masses
are smaller than $\var$. There is a solution ($n=1$) only when
$k=0$. Namely, a superpotential can be generated
only when $\var$ is smaller than any of the real bare masses,
and it is by a one-instanton effect.

\subsection{Pure Yang-Mills Theory}

Let us consider the case of pure $N=2$ $SU(2)$ Yang-Mills theory.
In view of the above analysis, a superpotential may be generated.
In fact, it was first shown in \cite{AHW} that it is indeed
generated. Using the dilute instanton gas approximation one finds
it has the form \cite{AHW}
\beq
W=\e^{-\phi}\,,
\label{pot}
\eeq
where $\phi$ is the chiral multiplet dual to the linear
multiplet whose lowest component is the complex scalar field
of the form $\phi=I+i\sigma$. Here, $I$ depends on $\var$
and approaches the classical lagrangian $\var/2e^2$ in the limit
$\var/e^2\to \infty$. Thus in the weak coupling region,
the Coulomb branch is completely lifted
and there is no supersymmetric vacuum.

\subsection{The Single Flavor Case}

Let us next consider the case with a single quark multiplet.
From the above selection rule (\ref{au1}), when the complex bare mass
is turned off $\mc=0$, a superpotential may be generated in the
region $\var<\mr$ but is never generated when $\var > \mr$.   
If we send $\mr$ to infinity, we expect to
recover the superpotential (\ref{pot}) of pure $N=2$
Yang-Mills theory. It implies that a superpotential is non-zero
in the region $\var <\mr$ and is zero when $\var>\mr$.
This seems peculiar because the superpotential is a holomorphic
function but a holomorphic function vanishing on one region
must be vanishing everywhere.
If the Coulomb branch is like the classical
half-cylinder (\ref{classmet}),
there must be a jump of the superpotential on the ``phase boundary''
$C=\{\var=\mr, 0\leq \sigma\leq 2\pi\}$
of real codimension one, which should not be the case.

As mentioned in section 2.6,
the only possible resolution to this puzzle
is that the Coulmb branch degenerates to have at least
two branches and the superpotential
is non-zero on one branch and is zero on the other.
As we now see, this is actually the case.
The ``phase boundary'' $C$ shrinks to a point.

Note that in the limit $\var,\mr\gg e^2$,
$|\var-\mr|\,\,\displaystyle{\mathop{<}_{\sim}}\,\, e^2$,
the off-diagonal pieces of the
$SU(2)$ vector multiplet (W-bosons and their superpartners)
decouple and
we can well approximate the system by
an $N=2$ $U(1)$ gauge theory with $N_f=1$ electron.
As discussed in the previous section, the Coulomb branch of the
latter system is determined at the one loop level and
the metric is given by (\ref{abel}) in which we identify
$r$ with $\var-\mr$. It
behaves in the limit $|\var-\mr|/e^2\to 0$ as
\beqa
ds^2&=&{1\over 4|\var-\mr|}d\var^2+|\var-\mr|d\sigma^2
\nonumber\\
&=&\left(d\sqrt{|\var-\mr|}\right)^2+\sqrt{|\var-\mr|}^2d\sigma^2\,
.
\label{nf1met}
\eeqa
From this expression we see that the ``phase boundary''
$C$ = $\{\var=\mr,0\leq\sigma\leq 2\pi\}$
has shrunk to a point and the Coulomb branch has
degenerated to have two branches corresponding to
$\var\leq \mr$ and $\var \geq \mr$.
The metric (\ref{nf1met}) on each branch is the flat
metric on the complex line with the coordinate
$x=\sqrt{\mr-\var}\e^{-i\sigma}$ for the branch
$\var\leq\mr$
and $y=\sqrt{\var-\mr}\e^{i\sigma}$ for $\var\geq\mr$.
Note that
this region of the Coulomb branch 
is simply described by
\beq
xy=0\,,
\label{xy0}
\eeq
as has already been seen in  \cite{dhoy}.
A Higgs branch is emanating from the singular point $x=y=0$.
Classically, it is described by the K\"ahler quotient
of $\C\oplus\C^{\vee}$ by the $U(1)$ action
(with vanishing FI parameter in this case), and is the
orbifold $\C/\Z_2$ of deficit angle $\pi$ at the origin
which is the intersection point with the Coulomb branch.

When the complex bare mass is turned on $\mc\ne 0$,
the description (\ref{xy0}) changes to
\beq
xy=\mc\,,
\eeq
and the singularity at $x=y=0$ is deformed to be smooth.
Due to the tree level superpotential $\mc \widetilde{Q}Q$
the Higgs branch is lifted.

Now, we complete the description of vacua in this weak coupling
region by taking into account the non-perturbative superpotential.
From the zero mode counting and by considerating the limit
$\mr\to\infty$, we expect in the case $\mc=0$ that the 
one-instanton configurations generate the superpotential
$W=\exp(-I-i\sigma)$ in the branch $\var\leq \mr$.
In the branch $\var\geq \mr$
the superpotential is not generated, $W=0$. In the
limit $\mc\to \infty$, we also expect to obtain the
superpotential (\ref{pot}).
The superpotential with this property is uniformly expressed
as
\beq
W=\e^{-\phi}
\eeq
where $\phi$ is a chiral superfield whose scalar component
is a holomorphic function on the Coulomb branch
such that ${\rm Im}\phi=\sigma$. In the particular region of 
interest,
$\phi$ is approximately the one
given in terms of $\var$ and $\sigma$
by (\ref{one})
applied to the case of $U(1)$ gauge theory with electrons
of charge $\pm 1$.
In other words, the superpotential has the lowest component
\beqa
W&=&\e^{-{\var\over 2e^2}-i\sigma}
\left(\frac{(-\var-\mr)+\sqrt{(-\var-\mr)^2
+|\mc|^2}}
{(\var-\mr)+\sqrt{(\var-\mr)^2
+|\mc|^2}}
\right)^{1/2}\nonumber\\
&\sim&\e^{-{\var\over 2e^2}-i\sigma}
\left(\frac{|\mc|^2}
{(\var-\mr)+\sqrt{(\var-\mr)^2
+|\mc|^2}}
\right)^{1/2}
\eeqa
Indeed, in the limit $\mc\to 0$,
$W=0$ for $\var>\mr$ and
$W=\sqrt{\mr-\var}\e^{-{\var\over 2e^2}-i\sigma}$
for $\var<\mr$. Also, in the limit $\mc\to\infty$,
it becomes $\e^{-{\var\over 2e^2}-i\sigma}$.
Note that near the singularity $\var=\mr$
of the $\mc=0$ model, the superpotential behaves in the branch
$\var\leq \mr$ as
\beq
W=\e^{-\phi}\sim \sqrt{\mr-\var}\e^{-i\sigma}= x\,.
\eeq
We can modify the definition of $x$ to be $x=\e^{-\phi}$
so that $W=x$ is exact for all values of $\mr$ and $\mc$.
When $\mc=0$,
since $\partial W/\partial x=1$ and $ds^2=dxd\overline{x}$
near the point $x=0$,
the branch $\var\leq\mr$ is completely lifted (except for possible
vacua in the strong coupling region which is not under control).
The branch $\var\geq \mr$ remains as supersymmetric vacua.
When the complex bare mass is turned on $\mc\ne 0$,
the Coulomb branch is described by $xy=\mc$, and as
$\partial W/\partial x=1$ and $\partial W/\partial y=-\mc y^{-2}$,
the whole Coulomb branch is lifted.

The Higgs branch, which is classically present for $\mc=0$,
may possibly be lifted by the generation of a superpotential.
Here we exclude this possibility.
In this situation, a superpotential can only be generated by
instantons.
Thus, it must be a chiral superfield
containing a power of $\e^{-i\sigma}$.
Namely, it must contain a positive power of $\e^{-\phi}$. 
However, $\e^{-\phi}$ vanishes at $\var=\mr$ where the
Higgs branch is emanating. This shows that there is no dynamical
generation of superpotential on the Higgs branch.
The metric on the Higgs branch
may be corrected by non-perturbative
effects, but the correction is very small in the region
$\var=\mr\gg e^2$.

It is a subtle issue whether the point $x=y=0$ intersecting with
the Higgs branch corresponds to a supersymmetric vacuum. The derivative
of the superpotential $W$ is vanishing in the directions of Higgs
branch and of the component $\var\geq\mr$ of the Coulomb branch
but is non-vanishing in the direction of the branch $\var\leq \mr$.
However, we should note that the lagrangian description in terms
of the variables we have used is not valid at this point. Thus,
there is no reason to exclude this point from the set of vacua.
If it is really a supersymmetric vacuum, it may correspond
to a non-trivial $N=2$ superconformal field theory.

\begin{figure}
\begin{center}
$\mbox{\epsfig{figure=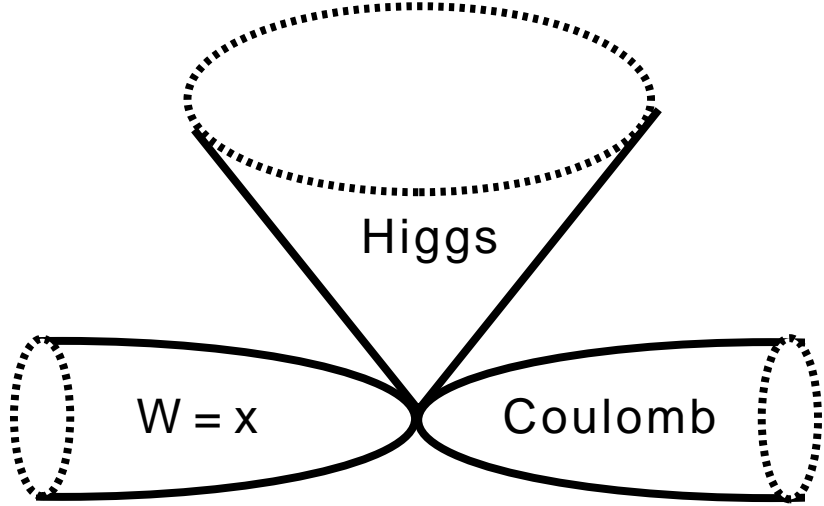}}$
\end{center}
\caption{The moduli space of vacua at weak coupling.}
\end{figure}

To summarize, the $SU(2)$ gauge theory with one quark multiplet
of large real bare mass $\mr\gg e^2$ and
vanishing $\mc=0$ has two branches in the weak coupling
region: one is a Higgs branch with a conical singularity and the other
is a Coulomb branch. They might intersect at one point, as in 
figure 2.
When $\mc$ is turned on, both the Higgs and Coulmb branches are
lifted completely. 

Notice that we did not take instanton corrections to the K\"ahler
potential into account. If such corrections exist and are e.g. of
the form $|e^{-\Phi}|^2$, with $\Phi$ the holomorphic field on the Coulomb
branch $r>m_r$, they may diverge as $r\downarrow m_r$ and change
the picture. In \cite{them} it was argued that the Coulomb and
Higgs branches merge. If that is the case they 
are connected by an exponentially small
neck that vanishes as $m_r \rightarrow \infty$, and figure~2
applies to this limit.  

\subsection{The Multi-Flavor Case}

For the case with $N_f>1$, the analysis is similar
and hence we present only the result. As noted previously,
we assume that all the real mass parameters are much larger than
$e^2$, $\mr^a\gg e^2$. The Coulomb branch degenerates at each point
$\var=\mr^a$ such that $\mc^a=0$, and from these points a Higgs branch 
emanates.
If the number of massless quark multiplets is one, the local 
structure
of the moduli space is of course the same as in the $N_f=1$ case.
If there are $N$ massless quarks
at a point, i.e., if $N$ quarks have the same real mass and 
vanishing
complex mass parameters, the Coulomb branch degenerates at the 
point
and locally looks like
two cones of the type $\C/\Z_N$ intersecting at the origin.
The Higgs branch
emanating from this point is the singular subspace of
$\C^{N^2}=\{x_{a,b}\}$
defined by $x_{a,b}x_{c,d}=x_{a,d}x_{c,b}$
which has a complex dimension $2N-1$.
The superpotential is uniformly
written as
\beq
W=\e^{-\phi}\sim\e^{-{\var\over 2e^2}-i\sigma}
\prod_{a=1}^{N_f}
\left(
\frac{|\mc^a|^2}{
\var-\mr^a+\sqrt{(\var-\mr^a)^2+|\mc^a|^2}}
\right)^{1/2}.
\eeq
Note that the superpotential is vanishing in the branches
in $\var\geq \mr^{a_0}$ where $\mr^{a_0}$ is the smallest $\mr^a$
such that $\mc^a=0$, and non-vanishing in the branch 
$\var\leq\mr^{a_0}$.
If all the complex bare masses are non-zero, then,
$W$ is nowhere vanishing.

Near the region ${\rm min}\{\mr^a\}\leq\var\leq{\rm max}\{\mr^a\}$,
the Coulomb branch is holomorphically embedded
in an ALE space, which is a resolution of the complex surface 
described by
\beq
xy=\prod_{a=1}^{N_f}\,\,(z-\mc^a).
\eeq
The parameter of resolution is encoded in $\mr^a$.
Roughly speaking,
when $\mc^a$ are very small, there are $N_f-1$
two-spheres whose intersection matrix is the Cartan matrix of
$A_{N_f-1}$ type, and the size of the $a$-th sphere is given by
the difference $\mr^a-\mr^{a+1}$ of the neighboring real masses.
The embedding of the Coulomb branch is defined by
\beq
z=0\,.
\eeq
The superpotential is expressed as
\beq
W=x\,.
\label{sup}
\eeq
If all the complex masses are turned on,
the Coulomb branch, being described by $xy=$non-zero constant,
has a good coordinate $x$, and it is lifted because
$\partial W/\partial x=1$.
It is possible to see that the superpotential (\ref{sup})
behaves as mentioned above as we turn off some complex mass
parameters.

\begin{figure}
\begin{center}
$\mbox{\epsfig{figure=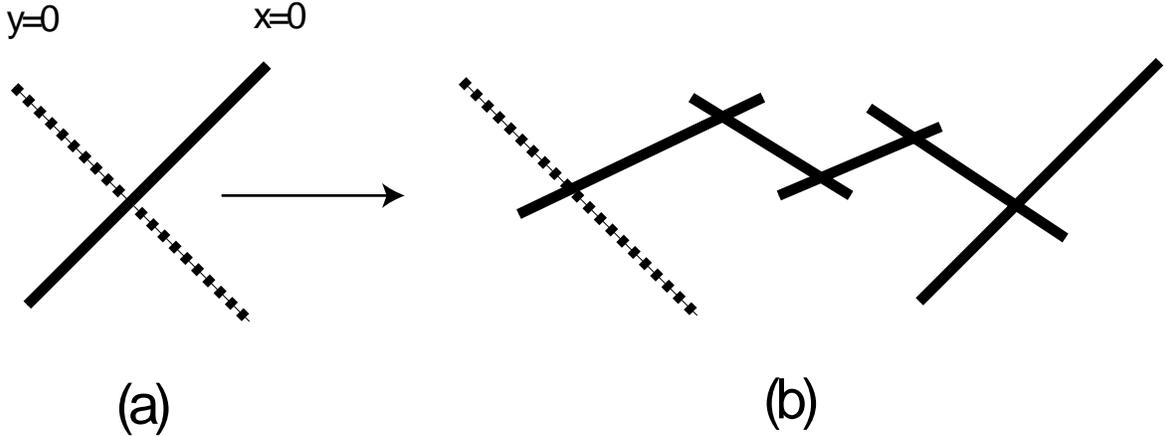}}$
\end{center}
\caption{Coulomb branches of $SU(2)$ gauge theories.}
\end{figure}

\noindent
As an ilustration, we present in figure 3 the Coulomb branch
for the case in which all the complex mass parameters are turned 
off.
Figure 3(a) describes the case in which all the real masses are the 
same
and 3(b) is for the case in which they are distinct.
The dashed lines are the branches that are lifted by the
superpotential (\ref{sup}), and the solid lines are the branches
that remain as supersymmetric vacua because (\ref{sup}) vanishes
identically on them. Intermediate branches
in fig 3(b) are ${\bf P}^1$'s whose sizes are
given by the difference of the neighboring
real mass parameters. In the scenario of \cite{them}, the Coulomb
branch that connects to the dashed line in figure~3(b) is
merged with a Higgs branch.

\section{Dynamics of $U(N_c)$ Gauge Theories}

In this section we will study  the superpotentials and the structure of the moduli 
spaces of vacua
for gauge groups $U(N_c)$  
and  $N_f$ pairs of chiral multiplets in the
fundamental representation.
The analysis and the structure that we will find generalizes that 
of the previous
section.
The case of $SU(N_c)$ gauge groups follows in a straightforward way 
from the analysis for 
$U(N_c)$ gauge groups and we will outline the differences.

As discussed previously,
 the bosonic part of the $N=2$ vector multiplet contains the three 
dimensional
gauge field $A_{\mu}$ and a real scalar $\varphi$ corresponding to 
the $A_4$ component of the
four dimensional gauge field. The terms $Tr[A_{\mu},A_{\nu}]^2$ and 
$Tr[A_{\mu},\varphi]^2$ in the $N=2$ action
imply that the Coulomb branch of the theory is parametrized by the 
vev's of the scalars
 $\varphi_i$ taking values in the Cartan sub-algebra of the gauge 
group
 and the the scalars $\sigma_i$ dual to the photons.
 The scalars in the quark multiplets parametrize the Higgs 
branch of the theory.

The metrics on the moduli spaces of vacua receive both loop and non 
perturbative
corrections. A superpotential can only be generated non 
perturbatively.
As discussed previously, one expects the non-perturbative effects 
to come
only from instantons which are monopoles in three dimenions.
Their magnetic charges correspond to the simple roots of the Lie 
algebra.  
The regions of weak coupling are defined by $|r_i - r_j| \gg e^2$.
In these regions the instanton factor 
is small and the instanton calculus is reliable.

\subsection{$U(3)$ Gauge Group}

In this subsection we will analyse  the $N=2$ supersymmetric gauge 
theories
with gauge group $U(3)$. 
We will consider first the case with no quarks $N_f=0$.

\subsubsection{Pure Yang-Mills Theory}

The classical metric on the Coulomb branch is
\beq
ds^2 =\sum_{i=1}^3\left( \frac{1}{4e^2} dr_i^2 + e^2 
d\sigma_i^2\right)
\comma
\eeq
 where
$\sigma_i$ are periodic with period $2\pi$. The classical moduli space of vacua
is that of three cylinders $\{r_i, \sigma_i\}$ modded out by 
by the action of the Weyl group.
 
This metric receives loop and instanton corrections.
In order to see whether the moduli space of vacua is lifted or not 
we have to check whether
a superpotential is generated.
The index formula for the fermionic zero modes in the background
of a monopole of a charge vector $(n_1,n_2)$
 takes the form
 \beq
  2(n_1+n_2) = 2
  \stop
  \eeq
 Therefore  only the fundamental 
monopoles
 with charge vectors $(1,0)$ and $(0,1)$
contribute to the superpotential.
The corresponding instanton terms are $\e^{-(\phi_1-\phi_2)}$ and    
$\e^{-(\phi_2-\phi_3)}$
respectively.
The superpotential reads
\beq
W = \e^{-(\phi_1-\phi_2)} +  \e^{-(\phi_2-\phi_3)}
\stop
\label{super}
\eeq
Thus, in the weak coupling region the Coulomb branch is completely 
lifted at the quantum
level.

\subsubsection{The Single Flavor Case }

Consider now adding one quark, namely
$N_f=1$.
Each quark generates two fermionic zero modes with opposite sign 
to the
gluino zero modes.
However when  the complex  mass parameter $m_c$ of the quark
 is diferent than zero the two
extra zero modes  are lifted. Thus the superpotential
generated in the case with  one quark and
$m_c \neq 0$
 is the same
as
that  with no quark multiplets (\ref{super}), and the 
classical moduli space
of vacua is lifted by instantons.

Consider now the case with $m_c = 0$. 
Without loss of generality we consider the region  $r_1 > r_2 
>r_3$.
The other regions can be analyzed similarly, or alternatively can 
be obtained by
the action of the Weyl group.
In order to analyze the possible monopoles contributing to the 
superpotential we have 
to distinguish between several regions depending on 
 the real mass parameter $m_r$ of the quark.

\noindent
(1) {\it $r_1 < m_r$ or $r_3 > m_r$}:
In this case 
the index  formula is the same as in the case $N_f=0$, namely the 
quark
does not contribute zero modes.
This implies that  in this region both the fundamental
monopoles with charge vectors
$(1,0)$ and $(0,1)$ contribute to the superpotential.

\noindent
(2) {\bf $r_1 > m_r > r_2 $}: In this case
the index formula for the fermionic zero modes in the background
of a monopole of charge vector $(n_1,n_2)$
 takes the form 
 \beq
 2(n_1+n_2) -2n_1= 2
 \stop
 \eeq
  Therefore only the fundamental monopole
 with charge vector  $(0,1)$
contributes to the superpotential.

\noindent
(3) {\bf $r_1  > r_2 > m_r > r_3 $}: In this case
the index formula for the fermionic zero modes in the background
of a monopole of a charge vector $(n_1,n_2)$
 takes the form 
 \beq
 2(n_1+n_2) -2n_2= 2
 \stop
 \eeq
  Therefore  only the fundamental monopole
 with charge vectors  $(1,0)$
contribute to the superpotential.
The  structure of monopole contributions to the superpotential is 
depicted in figure 4.

\begin{figure}
\begin{center}
$\mbox{\epsfig{figure=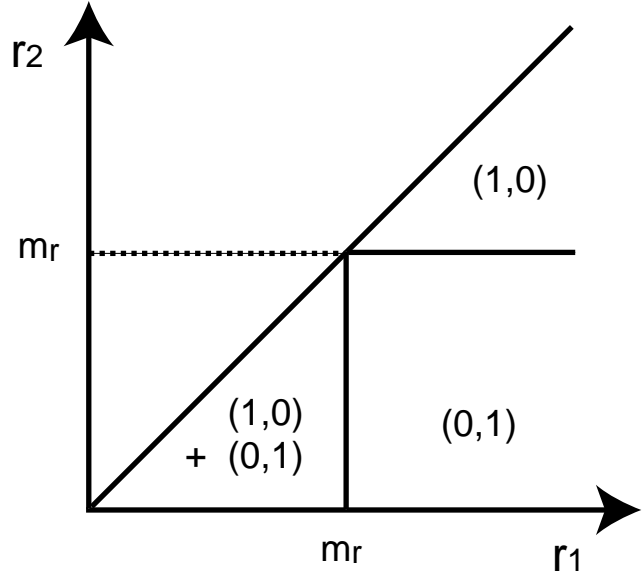}}$
\end{center}
\caption{Monopole contributions to the superpotential for $U(3)$ 
gauge group
with one quark
and  $m_c=0$.}
\end{figure}

This structure that we arrived at by analyzing
the fermionic zero modes in the monopole background
can also be seen from the one loop expression for $\phi$ 
(\ref{one})
in the weak coupling region $|r_i - r_j| >> e^2$.
Alternatively, the expression for
$\phi_i$ can be computed as
\beq
\phi_i = \int g_{ii} dr_i + i\sigma_i
\comma
\label{phi}
\eeq
where $g_{ij}$ is the one loop metric on the Columb branch in the 
weak coupling regions, which can be extracted from \cite{dhoo}.
In these regions the off diagonal terms of the metric can be 
neglected, and the diagonal
terms take the form
\beq
g_{ii} = \frac{1}{4}\left(\frac{1}{e^2} + \frac{1}
{\sqrt{(r_i-m_r)^2 + |m_c|^2}}\right)
\stop
\label{g}
\eeq

Using (\ref{phi}) and (\ref{g}) we get for 
the instanton factor $\e^{-\frac{\phi_i-\phi_{i+1}}{e^2}}$ 
\beq
\e^{-\frac{\phi_i-\phi_{i+1}}{e^2}} = 
\e^{-\frac{r_i-r_{i+1}}{2e^2} -i(\sigma_i-\sigma_{i+1})} 
\left(\frac{(r_{i+1} - m_r) + \sqrt{(r_{i+1}-m_r)^2 + |m_c|^2}}
{(r_i - m_r) + \sqrt{(r_i-m_r)^2 + |m_c|^2}}\right)^{\frac{1}{2}}
\stop
\label{instanton}
\eeq

When $m_c \neq 0$ the instanton factor
(\ref{instanton}) is non vanishing for both $i=1,2$
and indeed we get contributions to the 
superpotential from  the fundamental monopoles of charges
$(1,0)$ and $(0,1)$, as expected in this case.

When $m_c = 0$ we have to distinguish several  regions.
In the regions  $r_1 < m_r$ or $r_3 > m_r$  the instanton factor
(\ref{instanton}) is non vanishing 
for both $i=1,2$ and  we get contributions to the 
superpotential from both the $(1,0)$ and $(0,1)$ monopoles.
In the region  $r_1 >m_r > r_2 $ the instanton factor 
$\e^{-(\phi_1-\phi_2)}$ vanishes
and therfore only the $(0,1)$ monopole contributes, while
in the region $r_1  > r_2 > m_r > r_3 $ the instanton factor
$\e^{-(\phi_2-\phi_3)}$
vanishes and only the $(1,0)$ monopole contributes.
Thus, we see a complete agreement between the zero mode analysis 
and the one-loop
expression for the instanton factors.

As is clear from the analysis and from figure 1, in the weak 
coupling
regions a superpotential is generated everywhere
except at the region $r_1>m_r, r_2 = m_r$. 
Therefore the Coulomb branch is lifted everywhere except for that 
region. Classically, there is a Higgs branch emanating from this region.
Since the instanton factors corresponding to the fundamental monopole
vanish at this region one might expect no non-perturbatively 
generated superpotential on
the Higgs branch. However, the zero mode analysis fails in this region.
Far out in the Higgs branch the theory contains a gauge theory of the unbroken $U(2)$
gauge group with no matter, which we know has no supersymmetric vauca\footnote{We
would like to thank O. Aharony for suggesting this argument.}. Therefore,
the Higgs branch is lifted, due to charge $(1,1)$ monopoles, which are the
fundamental monopoles for the unbroken $U(2)$. We expect the same monopoles
to also lift the root of the Higgs branch.

Summarizing the structure of the moduli space of vacua at weak 
coupling:
The superpotential of the theory is given by (\ref{super}) with the instanton
factors of (\ref{instanton}).
At the classical level we have both Coulomb and Higgs branches, 
while at the quantum level both
the Coulomb and the Higgs branches are lifted.

Note that the diagonal $r_1=r_2$ in figure 4 corresponds to the strong coupling region 
of the 
theory where the instanton calculus is not reliable and therefore the methods that we 
use
in order to study the moduli space of vacua are not
valid.
However, as discussed previously, we can extrapolate the results to these region using
holomorphy and global symmetries provided
that the K\"ahler potential does not develop extra singularities, and conclude 
that all branches remain lifted.

\subsubsection{The Two-Flavor Case $N_f=2$}

The analysis of the theory with two quarks
is similar to that of the single quark case.
An interesting case is when the complex mass parameters of the two 
quarks vanish and
we get four extra fermionic zero modes.   
Consider the region $r_1 > m_r^1, m_r^2 > r_2 > r_3$.
The index formula for the fermionic zero modes in the background of 
a charge $(n_1,n_2)$ monopole
reads
\beq 
2(n_1+n_2) - 4n_1 = 2
\stop
\eeq
This suggests that in this region all monopoles of charges $(n,n+1)$ are 
allowed to contribute
to the superpotential. However, the other selection rule (\ref{selec2}) shows
that $n=0$ and that
only the fundamental monopole of 
charge $(0,1)$ contributes
to the superpotential.
In order to show that from field theory we can derive, 
 as in the single flavor
case, the instanton factor for the two flavor case 
using (\ref{phi}) and
the expression for the one loop metric. We get

\beq
\e^{-\frac{\phi_i-\phi_{i+1}}{e^2}} = 
\e^{-\frac{r_i-r_{i+1}}{2e^2} -i(\sigma_i-\sigma_{i+1})} 
\left( \prod_{a=1}^{2}
\frac{(r_{i+1} - m_r^a) + \sqrt{(r_{i+1}-m_r^a)^2 + 
|m_c^a|^2}}{(r_i - m_r^a) + 
\sqrt{(r_i-m_r^a)^2 + |m_c^a|^2}}\right)^{\frac{1}{2}}
\stop
\label{inst}
\eeq
Using (\ref{inst}) in the region $r_1 > m_r^1, m_r^2 > r_2 > r_3$, 
we see that the instanton factors
corresponding to monopoles of charges $(n,n+1)$ vanish unless 
$n=0$.

\begin{figure}
\begin{center}
$\mbox{\epsfig{figure=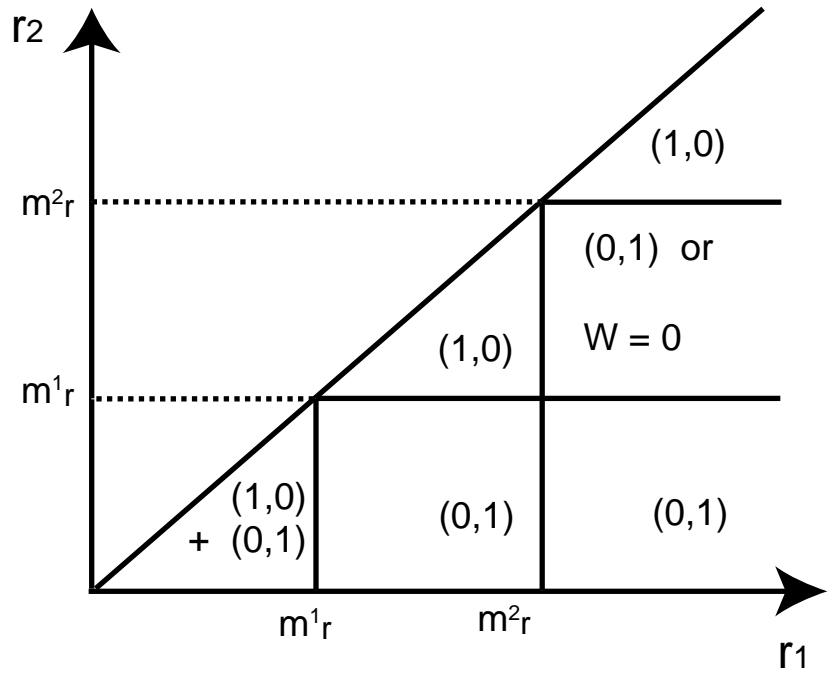}}$
\end{center}
\caption{Monopole contributions to the superpotential for $U(3)$ 
gauge group
with two quarks
and  $m_c=0$.}
\end{figure}

Performing the zero modes analysis as for the single flavor case we get 
the different monopole contributions to the superpotential as depicted in 
figure 5.
Note that in the region $r_1 > m_r^2 > r_2 > m_r^1$ there exist two 
possibilities
depending on $r_3$.
When $r_3 < m_r^1$ there is no generation of a superpotential on 
the Coulomb branch
by instantons.
When $r_3 > m_r^1$ the monopole with charge vector $(0,1)$ 
contributes and a superpotential
is generated. Classically, there are Higgs branches emanating
from the lines $r_1=m_r^1$, $r_1=m_r^2$,  $r_2=m_r^1$ and
$r_2=m_r^2$. All Higgs branches are lifted except the ones emanating from
the intervals that border $W=0$ in figure~5. The Higgs branches with
a non-trivial instanton factor are trivially lifted. To analyze
the ones with vanishing instanton factor one can go far out on the
Higgs branch and consider whether the effective field theory there
has supersymmetric vacua or not.

Summarizing the quantum picture: 
The superpotential of the theory is given by (\ref{super}) with the instanton
factors of (\ref{inst}).
Most of the Coulomb branch is 
lifted. 
In the region $r_1 > m_r^2 > r_2 > m_r^1$ with
$r_3 < m_r^1$ there is no generation of a superpotential on the 
Coulomb branch by instantons and there is a quantum flat direction.
Higgs branches emanate from the
regions $r_1 > m_r^2, r_2 = m_r^1$,
$r_1 > m_r^2, r_2 = m_r^2$ and 
$r_1 = m_r^2, m_r^2 > r_2 > m_r^1$.
In the picture of \cite{them}, these Higgs branches
merge with the Coulomb branch.

\subsubsection{The Multi-Flavor Case}

For general $N_f$ we have to distinguish more regions.
When the complex  mass parameter $m_c^a$ of the $a$'th quark is 
different than zero the two
extra zero modes that exist when the quark was massless
are  lifted. This case is then reduces to the situation with 
$N_f-1$ flavors.
A monopole with charge $(1,0)$ contributes only if  none of the real 
mass parameters
is in the interval $(r_1,r_2)$ while 
a monopole with charge $(0,1)$ contributes only if  none of the real 
mass parameters
is in the interval $(r_2,r_3)$.
Again this  structure can be seen from the expression for the 
instanton factor

\beq
\e^{-(\phi_i-\phi_{i+1})} = 
\e^{-\frac{r_i-r_{i+1}}{2e^2}-i(\sigma_i - \sigma_{i+1})}\left( \prod_{a =1}^{N_f}
\frac{(r_{i+1} - m_r^a) + \sqrt{(r_{i+1}-m_r^a)^2 + 
|m_c^a|^2}}{(r_i - m_r^a) + 
\sqrt{(r_i-m_r^a)^2 + |m_c^a|^2}}\right)^{\frac{1}{2}}
\stop
\label{instanton1}
\eeq

In analogy to figure 4 and figure 5, we summmarize here
the  monopole  contributions to the superpotential in the different regions on
the Coulomb branch 
by
\beq
\sum_{i=1}^{2} \prod_{a, m_c^a\neq 0}^{N_f}\Theta 
((r_i-m_r^a)(r_{i+1}-m_r^a)) 
R_i
\comma
\label{theta}
\eeq
where $\Theta$ is the step function, and $R_i$ are the charge vectors
 of the fundamental
monopoles, $R_1=(1,0), R_2=(0,1)$.

Summarizing the quantum picture: 
The superpotential of the theory is given
 by (\ref{super}) with the instanton
factors of (\ref{instanton1}).
Most of the Coulomb branch is 
lifted. There are regions
in the Coulomb branch where a superpotential is not generated, as 
can be seen from
(\ref{theta}).
In addition, there are regions where
we pass between zero and non-zero superpotentials, from
which Higgs branches emanate. In case some of the real
masses are equal, there will also be a Higgs branch
if we pass a line where more than one quark becomes
massless, but only if no monopole contributes
to both sides of the region it separates. 
These lines may 
correspond to non-trivial $N=2$ 
superconformal field
theories.

\subsection{$U(N_c)$ Gauge Group}
The analysis for general $N_c$ follows the same lines
as that of $N_c=3$ in the previous
section.

\subsubsection{Pure Yang-Mills Theory}

The index formula for the fermionic zero modes in the background
of a monopole with charge vector $(n_1,n_2,...,n_{N_c-1})$
takes the form 
\beq
2\sum_{i=1}^{N_c-1} n_i =2
\stop
\eeq
Therefore  only the fundamental monopoles
with charge vectors  $R_i = (0,...,1...,0)$ (a $1$ in the 
$i$'th entry)
contribute to the superpotential.
The corresponding instanton terms are $\e^{-(\phi_i-\phi_{i+1})}$.
The superpotential reads
\beq
W = \sum_{i=1}^{N_c-1}\e^{-(\phi_i-\phi_{i+1})} 
\stop
\label{super1}
\eeq
This form of the superpotential has been derived by considering 
M-theory on a 4-fold
in \cite{KV} and using open D-string instantons in \cite{dhoy}.
Here we get the same result from a field theory viewpoint.
We thus see that in the weak coupling regions the Coulomb branch is 
completely lifted at the quantum
level.

\subsubsection{The Multi-Flavor Case}

It is now clear how to generalize to $U(N_c)$ with $N_f$ flavors.
When the complex  mass parameter $m_c^a$ of the $a$'th quark is
different than zero the two
extra zero modes that exist when the quark was massless
are lifted. This case then reduces to the case with
$N_f-1$ flavors.
A monopole with charge vector $R_i = (0,...,1...,0)$ 
contributes only if  none of the real mass parameters
is in the region $(r_i,r_{i+1})$.

Performing the zero modes analysis we see that
 the monopole contributions to the superpotential in the different
 regions of the Coulomb branch can be summarized by

\beq
\sum_{i=1}^{N_c-1} \prod_{a, m_c^a\neq 0}^{N_f}\Theta 
((r_i-m_r^a)(r_{i+1}-m_r^a))R_i 
\label{for}
\stop
\eeq

As in the previous cases, the structure of  monopole contributions to the 
superpotential
(\ref{for}) can be seen from the one loop expression for 
$\phi_i$ 
(\ref{instanton1}).
For $N_f<N_c-1$ the Coulomb branch is lifted completely, by simple
zero-mode analysis. Classically, there are Higgs branches whenever
a certain number of quarks, say $p$, become massless. However,
far out on the Higgs branch the theory looks like an $N_c-1$
gauge theory with $p-1$ flavors, and by induction one sees
that in the quantum theory all Higgs branches are lifted as well.
For $N_f \geq N_c-1$, there can be Coulomb and Higgs branches
that remain in the quantum theory. 
Regions in the classical 
Coulomb branch where a superpotential is not generated can be read off
from (\ref{for}). The conditions for a Higgs branch to emanate
are similar to those discussed at the end of section~4.1.4.

\subsection{$SU(N_c)$ Gauge Group}

The analysis for $SU(N_c)$ gauge groups is similar to that of the 
$U(N_c)$ case.
The difference is that now the coordinates $\{r_i\}$ 
are subject to the restriction $\sum_{i=1}^{N_c} r_i=0$.
The metric that we have to integrate in order to get the expression 
for the
instanton factors is that of $SU(N_c)$ which is obtained from that 
of the $U(N_c)$ metric
by restricting to  the $SU(N_c)$ part. Alternatively, we can use formula 
(\ref{one}).
In order to compare the $SU(N_c)$ case to that of $U(N_c)$ which
we studied in detail in the previous section
consider the region $r_1 > r_2 >...> r_{N_c}=- \sum_{i=1}^{N_c-1} r_i$.

The  difference between the $U(N_c)$ and $SU(N_c)$ cases is 
that while for $U(N_c)$
we can take all $\{r_i\}$  to be positive this cannot be done for 
$SU(N_c)$.
This implies, for instance, that it is not possible in this region for $SU(N_c)$ 
to have $r_{N_c} > m_r^a$ if $m_r^a > 0$
where $m_r^a$ is the real mass parameter of the $a$'th quark.
Therefore such ranges of parameters in which there are instanton 
terms contributing
to the superpotential are excluded in the $SU(N_c)$ case.
In order to illustrate this consider for 
instance the $SU(3)$ with two flavors case.
In figure 5 we noted that for $U(3)$ with two flavors in the 
region $r_1 > m_r^2 > r_2 > m_r^1$ there were two possibilities
depending on $r_3$.
When $r_3 < m_r^1$ there was no generation of a superpotential on 
the Coulomb branch
by instantons.
When $r_3 > m_r^1$ the monopole with charge vector $(0,1)$ 
contributed and a superpotential
was generated.
This does not depend whether $m_r^1$ is positive or negative.
Since for $SU(3)$ the  case $r_3 > m_r^1> 0$  is excluded, there is 
only one possibility 
which is that the superpotential vanishes.

The superpotential for gauge group $SU(N_c)$ with $N_f$ flavors is given by
(\ref{super1}) with the instanton factors
derived from (\ref{one}). The instanton factors are 
identical to those in (\ref{instanton1})
for $i \neq N_c -1$, while  for $i =  N_c -1$ we get
\begin{eqnarray}
\lefteqn{\e^{-(\phi_{N_c-1}-\phi_{N_c})} = 
\e^{-\frac{r_{N_c-1}+ \sum_{i=1}^{N_c-1} r_i}{2e^2} +i(\sigma_{N_c-1}
+ \sum_{i=1}^{N_c-1}\sigma_i)}\times}
\label{instanton2}\\
&&\left( 
\prod_{a=1}^{N_f}
\frac{\left( -\sum_{i=1}^{N_c-1} r_i- m_r^a + 
  \sqrt{( -\sum_{i=1}^{N_c-1} r_i- m_r^a)^2 + |m_c^a|^2}\right)^{N_c}}
 {\left(r_{N_c-1} - m_r^a + 
   \sqrt{(r_{N_c-1}-m_r^a)^2 + |m_c^a|^2}\right)
  \prod_{i=1}^{N_c-1}\left( r_i - m_r^a + 
           \sqrt{(r_i-m_r^a)^2 + |m_c^a|^2}\right) }
\right)^{\frac{1}{2}}\nonumber
\stop
\end{eqnarray}

\subsection{D-brane picture}

A useful alternative viewpoint on the above results is to realize
the gauge theories in string theory using intersecting branes.
The $N=2$ gauge theories in three dimensions can be realized on the 
worldvolume of 
a D3 brane \cite{dhoy} following the construction of \cite{EGK,HW}. 
In order
to realize an $N=2$ gauge theory with $U(N_c)$ gauge group
and $N_f$ flavors, one uses  NS and NS${}^{\prime}$ 5-branes
with worldvolume coordinates $(x^0x^1x^2x^3x^4x^5)$ and 
$(x^0x^1x^2x^3x^8x^9)$
respectively, $N_f$ D5 branes 
with worldvolume coordinates $(x^0x^1x^2x^7x^8x^9)$ and
$N_c$  D3 branes with worldvolume coordinates $(x^0x^1x^2x^6)$.
Instanton corrections to the superpotential
of the three dimensional gauge theory in the 
$(x^0x^1x^2)$-direction arise
from open D-string instantons
corresponding to D-strings stretching between D3 branes.
In \cite{dhoy} the superpotential for $N=2$ $U(N_c)$ Yang-Mills 
theory
has been derived using these open D-string instantons.

Here we want to include matter in the brane framework.
Matter multiplets in this picture arise from the open strings 
stretching
between the D5 branes and the D3 branes.
We already saw in the field theory analysis 
that quarks can introduce extra fermionic zero modes.
In the brane picture these zero modes are localized at the 
intersection points between the D5 branes and the open D-string 
worldsheet,
and arise from the open fundamental strings stretching
between the D5 brane and the D-string worldsheet.
Each D5 brane intersecting a D-string worlsheet gives rise in this 
way
to two fermionic zero
modes.
These two fermionic zero modes are the same as the ones we get from 
one flavor
in the field theory description when the complex mass vanishes, 
$m_c=0$.

The region in the
Coulomb branch where we should take these zero modes into account
depends on the point of intersection with the D5 brane.
In figure 4 we see a configuration where the D5 brane intersects 
the D-string worldsheet
in the region between $r_i$ and $r_{i+1}$.
Since the position of the D5 brane in the $x^3$ direction 
corresponds to the real mass
parameter of the corresponding matter multiplet \cite{dhoy}, this
corresponds to
 the region  $r_i > m_r > r_{i+1}$ in the analysis of the previous sections.
 Thus, the term $\e^{-\frac{\phi_i-\phi_{i+1}}{e^2}}$
will not contribute to the superpotential in this case, while 
the other instanton factors will.
It is easy to see that  the fermionic zero modes analysis that we 
make in the field
theory framework is identical to that in the branes language. The 
dictionary between the
field theory and the brane picture is that
the location of the extra fermionic zero modes coming from the D5 
brane
intersecting the D-string
worlsheet correspond to the location of the real mass parameters in 
the different
$\{r_i\}$
regions.

The complex mass parameter of a quark is represented in the brane 
picture
by the position of the corresponding D5 brane in $(x^4,x^5)$.
Thus, setting  $m_c \neq 0$ implies in the brane picture that the 
D5 brane does
not intersect the D-string worlsheet and therefore the two extra 
fermionic
zero modes which were localized at the intersetion point do not 
exist.
This provides further evidence (besides the consideration of the
limit $\mc\to\infty$) for our claim that there is no fermion
zero mode from the matter chiral multiplet in the case $\mc\ne 0$.
 \begin{figure}
\begin{center}
$\mbox{\epsfig{figure=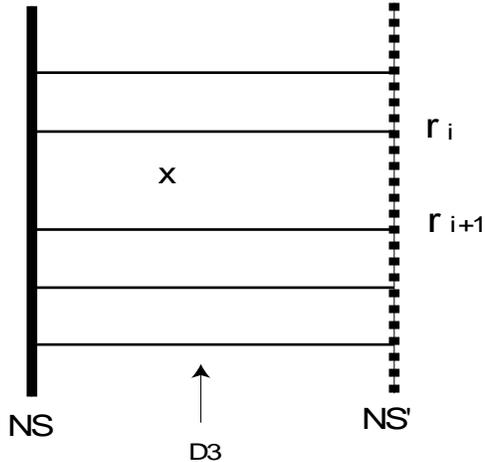}}$
\end{center}
\caption{D-brane picture.}
\end{figure}

\section{General non-abelian gauge theories}

In the previous sections we have shown that for gauge groups
$U(N_c)$ and $SU(N_c)$ with matter in the fundamental
representation  the 
non-perturbatively generated superpotential on the Coulomb branch 
is, at weak coupling, always of the form
\be \label{j14}
W=\sum_i \e^{-\alpha_i \cdot \Phi}
\ee
where the sum is over the simple roots of the Lie algebra, and the
result applies to the Weyl chamber where $\alpha_i \cdot <\Phi> 
>0$.
In this section we will argue that this result is in fact
valid for any gauge theory with arbitrary matter. To do this,
we will again examine the behavior of $W$ as we send the complex
masses $m_c\rightarrow 0$, and verify that whether $W$ vanishes
or not is precisely in agreement with the Callias index theorem.

Consider a quark transforming under a representation $R$
of the gauge group, with weights $w_{p}$, real mass $m_r$ and
complex mass $m_c$. According to the general result in (\ref{j11})
an exponential $\exp (-g \cdot \Phi)$, with $g$ an arbitrary
magnetic charge vector, contains, 
due to this quark, a factor
\be \label{gurb}
\e^{-g \cdot \Phi} \sim \e^{-g \cdot (\frac{r}{2e^2} + i\sigma)} 
\prod_p \left[ 
 (w_p \cdot r -m_r) +
  \sqrt{ (w_p \cdot r -m_r)^2 + |m_c|^2 } \right]^{-g\cdot w_p/2} .
\ee 
As we send $|m_c|\rightarrow 0$, the exponential behaves as
\bea
 \e^{-g \cdot \Phi} & \sim & 
 \prod_p \left[ 
 (w_p \cdot r -m_r)^{{\rm sign}(w_p \cdot r -m_r)} |m_c|^{1-
 {\rm sign}(w_p \cdot r -m_r) } \right]^{-g\cdot w_p/2} 
 \nonu
 & \sim & |m_c|^{N/2}
\eea
where
\be
N=\sum_p \frac{(w_p \cdot r -m_r)}{|w_p \cdot r -m_r|} (w_p \cdot 
g).
\ee
According to the results in the appendix, $N$ is precisely equal
to the number of zero modes of the fermions of the quark multiplet
in a monopole background with charge $g$ and scalar field vev
equal to $r$. In hindsight, this is what we would expect. The
one-loop corrections in $\Phi$ come from a one-loop determinant
and are proportional to the product of the masses of all massive
particles that are integrated out. By introducing a complex mass
$m_c$ we give a mass proportional to $m_c$ to the fermion zero
modes. As we take $m_c \rightarrow 0$, the one-loop determinant 
vanishes as $|m_c|^{N/2}$, where $N$ is the number of zero modes.

In order to have a finite contribution to the
superpotential as we send all $|m_c|$ to
zero, we need that there are no zero modes for the quarks
at all in that particular monopole background. This is in
complete agreement with the index theory calculation from which
we obtained that if $m_c=0$ only monopoles 
that satisfy
\be
2\sum_{i=1}^{{\rm rank}\, G} n_i \pm N = 2
\ee
can contribute. This identity can only be true 
in a background corresponding to a fundamental
monopole. These are obtained by embedding the standard 
$SU(2)$ monopole in an $SU(2)$ subgroup of the gauge group 
corresponding
to a simple root. They give rise to the superpotential
in (\ref{j14}), and we conclude that this is the only possible
consistent superpotential.

To summarize, the superpotential on the
Coulomb branch of $N=2$ gauge
theory in three dimensions has the form (\ref{j14}), where
$\Phi$ is given by (\ref{gurb}) in the one-loop approximation.

\section*{Acknowledgements}

We would like to thank O. Aharony for useful discussions.
This work is supported in part by 
NSF grant PHY-951497 and DOE grant DE-AC03-76SF00098. JdB is a 
fellow of
the Miller Institute for Basic Research in Science. 

\appendix{Callias index}

One of the main tools in the analysis in the paper was
to use the number of fermionic zero modes of matter
multiplets in the background of a monopole. 
We have seen how the number of zero modes
naturally appears in one-loop calculations. In his appendix we want
to show how one computes 
 the number of zero modes using the Callias index theorem.
We consider the case where we are on a point in the Coulomb branch 
were 
the gauge symmetry is maximally broken and there are no massless 
matter 
fields. If either of these two conditions is not satisfied, there 
is no
general result regarding the number of zero modes, although some
special cases, like spherically symmetric monopoles \cite{wein1},
can still be treated. 

The bosonic part of a pure $N=2$ gauge theory in three dimensions 
is
described by (taking $e=1$)
\be \label{jj1}
S= \int d^3 x \tr (\frac{1}{4} F_{ij} F^{ij} + \frac{1}{2} (D_i 
\varphi)^2).
\ee
where $\varphi$ is a scalar field transforming under the adjoint
representation.
Let $\varphi_0$ denote its vacuum expectation value. By assumption,
it is such that it breaks the gauge group completely to its maximal
abelian subgroup, and in addition all matter fields should be 
massive.
We can rewrite the action (\ref{jj1}) as
\be \label{jj2}
S=\int \tr ( \frac{1}{4} (F+\ast D\varphi) \wedge \ast (F+\ast 
D\varphi) +
 \frac{1}{4} (F-\ast D\varphi) \wedge \ast (F-\ast D\varphi)).
\ee
BPS monopoles are field configurations satisfying $F=\ast 
D\varphi$. 
For such configurations, the action in (\ref{jj2}) is equal to
$S=\int \tr(F \wedge D \varphi) = \int d\tr((F \varphi))$. This
can be rewritten as an integral over a two-sphere at infinity
of $\tr(F\varphi)$. For large radius, and in a fixed
direction,  $F$ behaves as $(\ast 
F)_i\sim 
\frac{r_i}{r^3} g_0$, and $\varphi$ is equal to $\varphi_0$, so 
that the
action for a BPS monopole configuration is $S=\tr(\varphi_0 g_0)$. 
Since $[F_{ij},\varphi]=D_i(D_j \varphi)-D_j (D_i \varphi) $ 
vanishes at
large distances, we can simultaneously diagonalize $\varphi_0$ and 
$g_0$,
and assume both have values in the Cartan subalgebra. Furthermore,
the quantization condition for the magnetic field says that
$\exp(2\pi g_0)$ should be the identity element in the group, and
therefore $g_0$ should be an integral linear combination of the
dual simple roots of the group under consideration,
\be \label{jj3} 
g_0=\sum n_i \alpha_i^{\ast}.
\ee

Now, consider some quark transforming in a representation 
$R$
of the group $G$. What we are interested in is the number of zero
modes of the Dirac operator 
\be \label{jj4}
D=  \ssl D + \varphi
\ee
acting on spinors in the representation $R$. In case $\varphi$ has 
no
zero eigenvalues in the representation $R$ (i.e. there is no
massless matter), the Callias index theorem \cite{callias} can
be used to compute the index of the Dirac operator. The theorem
states that the index is proportional to the integral over a 
two-sphere
at infinity of the two-form $UdUdU$, where $U=\varphi/|\varphi|$. 
However,
for gauge groups larger than $SU(2)$ this is a very complicated 
calculation, and it will be easier to follow the calculation in
\cite{wein1,wein2}. 

One may now compute that (using gamma matrices $\gamma_i = 
\sigma_i$)
\be 
D^{\dagger} D = -\ssl D \ssl D - \ssl D \varphi + \varphi^2  
= -D_i D_i  + i ( -(\ast F)_k - D_k \varphi) \sigma_k + \varphi^2 
= -D_i D_i - 2 \ssl D \varphi + \varphi^2
\ee
and 
\be 
D D^{\dagger} = -\ssl D \ssl D + \ssl D \varphi + \varphi^2  
= -D_i D_i  + i ( -(\ast F)_k + D_k \varphi) \sigma_k + \varphi^2
= -D_i D_i + \varphi^2
\ee
By hermiticity of $\varphi$, the second operator is positive 
definite,
showing that the number of zero modes can be written as
\be \label{jj5}
N= \lim_{M^2\rightarrow 0} \tr \left( \frac{M^2}{D^{\dagger} D + 
M^2} \right)  - 
\tr \left( \frac{M^2}{D D^{\dagger}  + M^2} \right).
\ee

The normalizable zero modes of $D$ contribute $1$ to $N$. However, 
there can be
a problem with (\ref{jj5}) if the continuous spectrum has a 
singular behavior
at zero, in which case there can be additional contributions to $N$ 
\cite{wein1,wein2}. In the case at hand, the long range behavior of $\varphi$ 
and $A_i$ in a fixed direction is
\be \varphi=\varphi_0 - \frac{1}{4\pi r} g_0 + {\cal 
O}(\frac{1}{r^2}),
A_i = \omega_i g_0 + {\cal O}(\frac{1}{r^2}),
\ee
where $\omega$ is the standard Dirac monopole of unit charge on 
$S^2$. Consider a fermion corresponding to a vector in the 
representation $R$
with weight $w$. If $w \cdot \varphi_0\neq 0$, the fermion decays 
exponentially
if it has an eigenvalue close to zero, and there are no problems 
with the
continuous spectrum. However, if $w \cdot \varphi_0 =0$, the 
fermion can see
an effective $1/r^2$ potential due to the terms of order $1/r$ in 
$\varphi$ and $A_i$. In such a potential, the spectrum can be 
singular, leading
to an incorrect result for the number of zero modes if one uses 
(\ref{jj5}).
However, if in addition $w\cdot g_0=0$, the fermion does not see 
the $1/r$
terms and again there is no problem with the continuous spectrum. 
To summarize,
the calculation of the number of zero modes using (\ref{jj5}) is 
reliable
if for all weights (i) $w \cdot \varphi_0 \neq 0$ or (ii) $w \cdot 
\varphi_0 = 
w \cdot g_0 = 0$. 

To continue, we introduce Dirac matrices 
\be 
\gamma_i = \left( \begin{array}{cc} 0 & i\sigma_i \\ -i \sigma_i & 
0 \end{array}  \right), \quad
\gamma_4 = \left( \begin{array}{cc} 0 & 1 \\ 1 & 0 \end{array}  
\right), \quad
\gamma_5 = \left( \begin{array}{cc} 1 & 0 \\ 0 & -1 \end{array}  
\right).
\ee
and the operator ${\cal D}=\gamma_i D_i - i \gamma_4 \varphi$. This 
is the Dirac operator in
the four-dimensional $N=1$ theory whose dimensional reduction 
yields the $d=3,N=2$
theory under consideration. Using these gamma matrices, we can 
rewrite
\be \label{jj6} 
N = \lim_{M^2\rightarrow 0} \tr \left( \gamma_5 \frac{M^2}{ - {\cal 
D}^2 + M^2} \right)
\ee
Next, we observe that we can write
\be \label{jj7}
N = \int d^3 x \partial_i J_i = \int_{S^2_{\infty}} \hat{r}_i J_i 
d^2 x
\ee
where 
\be 
J_i = \frac{-1}{2} \tr <x| \gamma_5 \gamma_i {\cal D} \frac{1}{ - 
{\cal D}^2 + M^2} |x>,
\ee
and $\hat{r}_i$ is a unit vector on the two-sphere.
We can further manipulate this expressions using the fact
that 
\be
-{\cal D}^2 = -D_j^2 + \varphi^2 + M^2 + G
\ee
where $G$ is 
\be
G = \left( \begin{array}{cc} -2 \ssl D \varphi & 0 \\ 0 & 0 
\end{array} \right).
\ee
and $G$ decays as $1/r^2$ for large $r$. If we now make an 
expansion
of $(-{\cal D}^2 + M^2)^{-1}$ as a power series in $G$, then only 
the
zeroeth and first order terms in $G$ can contribute to (\ref{jj7}), 
the
others decay too fast for that. The term independent of $G$ 
contains
$\tr(\gamma_5 \gamma_i \gamma_{\mu})$, $\mu=1\ldots 4$, which 
vanishes,
the term linear in $G$ contains the trace over gamma matrices
$\tr(\gamma_5 \gamma_i \gamma_{\mu} G)$. If $\mu\neq 4$ one gets a 
term
proportional to $\epsilon_{ijk} \hat{r}_i \hat{r}_j$, which 
vanishes.
Therefore, the only relevant term remaining is, at large $r$,
\be
J_i \sim \frac{\hat{r}_i}{r^2} \tr <x| \varphi_0 \frac{1}{-D_j^2 
+\varphi_0^2+M^2} g_0
\frac{1}{-D_j^2 +\varphi_0^2+M^2} |x>
\ee
The operator we trace is diagonal if we choose the weight basis for 
$R$, as
both $\varphi_0$ and $g_0$ can be chosen to lie in the Cartan 
subalgebra. This yields
\be
J_i \sim \frac{\hat{r}_i}{r^2} 
\sum_w (w \cdot \varphi_0) (w \cdot g_0) <x | 
\frac{1}{(-D_j^2 + (w\cdot \varphi_0)^2 + M^2)^2} |x>
\ee
The last factor is easily worked out in momentum space. Putting 
everything and the
correct normalizations together we finally obtain
\be
N=\lim_{M^2 \rightarrow 0 } \sum_w \frac{(w \cdot \varphi_0) 
(w \cdot g_0)}{((w \cdot \varphi_0)^2 + M^2)^{1/2}}
\ee
or equivalently
\be
N=\frac{1}{2} \sum_{w,(w\cdot \varphi_0)\neq 0} ({\rm sign}(w \cdot 
\varphi_0)) (w\cdot g_0).
\ee
This result can easily be extended to the case where there is a 
real mass $m$ for the
fermions, leading to the result
\be
N=\frac{1}{2} \sum_{w,(w\cdot \varphi_0)\neq m} ({\rm sign}((w 
\cdot \varphi_0)-m)) (w\cdot g_0).
\ee

As an example, consider a fermion in the adjoint representation of 
SU(k), and
take $\varphi_0={\rm diag}(\varphi_1,\ldots,\varphi_k)$, with 
$\varphi_i > \varphi_{i+1}$,
in which case $g_0={\rm diag}(n_1,n_2-n_1,\ldots,-n_{k-1})$ with 
the $n_i$ nonnegative. 
The nonzero weights have one entry $+1$ and one entry $-1$ and all 
others equal
to zero. This yields then 
\be
N_{adj}= \sum_{j>l} ((n_{j+1}-n_j)-(n_{l+1}-n_l)) = 2 \sum_i n_i,
\ee
the known result. 

Next, consider a fermion in the fundamental representation, with 
real mass $m$.
The weights can be taken as having one entry equal to one and all 
others equal to zero. 
Let t be the index such that $\varphi_t -m> 0 > \varphi_{t+1}-m$. 
If there is no such index $t$, then $N=0$, otherwise
\be
N_{fund}=n_t.
\ee
A quark multiplet contains two fermions, and has therefore $2n_t$ 
zero modes. Notice that the number of zero modes jumps as we
vary the $\varphi_i$.

Although we have not computed the index in the presence of a 
complex mass, the
results of the one-loop calculations strongly suggest that there 
are no 
zero modes left as soon as a complex mass parameter is turned on. 
It should be
quite easy to verify this in the above framework as well.


\newpage
 

 \begin{thebibliography}{99}

\small
\parskip=0pt plus 2pt
\bibitem{AHW} I. Affleck, J. Harvey and E. Witten,
``Instantons and (Super-) symmetry Breaking in (2+1) 
Dimensions,''
\np206,82,413.
\bibitem{KV} S.~Katz and C.~Vafa, ``Geometric Engineering
of $N=1$ Quantum Field Theories,'' hep-th 9611090.
\bibitem{wi} E. Witten, ``Non-Perturbative Superpotentials in
 String Theory'', hep-th 9604030, \np474,96,343.
\bibitem{dhoy} J.~de Boer, K.~Hori, Y.~Oz and Z.~Yin, 
``Branes and Mirror Symmetry in $N=2$ Supersymmetric Gauge Theories 
in Three Dimensions,'' 
 hep-th 9702154.
\bibitem{hklr} N.~J.~Hitchin, A.~Karlhede, U.~Lindstrom and 
M.~Rocek,
``HyperK\"ahler Metrics and Supesymmetry,'' \cmp117,88,569. 
\bibitem{intsei} K.~Intriligator and N.~Seiberg,
``Mirror Symmetry in Three Dimensional Gauge Theories,'' hep-th 
9607207.
\bibitem{dhoo} J. de Boer, K. Hori, H. Ooguri and Y. Oz, 
``Mirror Symmetry in Three-Dimensional 
Gauge Theories, Quivers and D-branes,'' hep-th 9611063.
\bibitem{dhooy} J. de Boer, K. Hori, H. Ooguri, Y. Oz and Z.~Yin, 
``Mirror Symmetry in Three-Dimensional 
Gauge Theories, $SL(2,Z)$  and D-brane Moduli Space,'' hep-th 
9612131.
\bibitem{EGK} S. Elitzur, A. Giveon and D. Kutasov,
``Branes and $N=1$ Duality in String Theory'', 
hep-th/9702014.
\bibitem{HW} A. Hanany and E. Witten,
``Type IIB Superstrings, BPS Monopoles, And 
Three-Dimensional
Gauge Dynamics'', hep-th/9611230.
\bibitem{wein1} E. Weinberg, 
``Fundamental Monopoles and Multi-Monopole Solutions for
 Arbitrary Simple Gauge Groups'', Nucl. Phys. {\bf B167} (1980) 500.
\bibitem{wein2} E. Weinberg, ``Fundamental Monopoles in Theories
with Arbitrary Symmetry Breaking'', \np203,82,445.
\bibitem{callias} C.J. Callias, 
 ``Index Theorems on Open Spaces,''
Commun. Math. Phys. {\bf 62} (1978) 213.
\bibitem{them} O. Aharony, A. Hanany, K. Intriligator, N. Seiberg and M.J. Strassler,
``Aspects of $N=2$ Supersymmetric Gauge Theories in Three Dimensions'', hep-th 9703110.
\end{thebibliography}
\end{document}